\documentclass[a4paper,11pt,preprintnumbers]{article}
\pdfoutput=1

\usepackage{amssymb,mathrsfs,amsmath}
\usepackage{a4wide}
\usepackage{color,xcolor}
\usepackage{slashed,soul}
\usepackage{graphicx}
\usepackage{amsfonts}
\usepackage{cancel}
\usepackage{lscape}
\def\linkcolor{cyan!70!black}

\usepackage[
colorlinks=true
,urlcolor=\linkcolor
,anchorcolor=\linkcolor
,citecolor=\linkcolor
,filecolor=\linkcolor
,linkcolor=\linkcolor
,menucolor=\linkcolor
,linktocpage=true
,pdfproducer=medialab
,pdfa=true
]{hyperref}
\usepackage{amsthm}
\usepackage{booktabs}
\usepackage{array}
\usepackage{rotating}
\usepackage[numbers, sort&compress]{natbib}
\usepackage{multirow}
\usepackage{float}
\usepackage[utf8]{inputenc}
\usepackage[T1]{fontenc}
\usepackage{appendix}
\usepackage{epsfig}
\usepackage{orcidlink}
\usepackage{comment}
\usepackage{arydshln}
\usepackage{xfrac}
\usepackage{feynmp-auto}
\usepackage{siunitx}
\usepackage{physics}
\usepackage[normalem]{ulem}

\usepackage{scalerel}

\newcommand{\beq}{\begin{equation}} 
\newcommand{\eeq}{\end{equation}} 
\newcommand{\ba}{\begin{array}}  
\newcommand{\ea}{\end{array}} 
\newcommand{\bea}{\begin{eqnarray}}  
\newcommand{\eea}{\end{eqnarray} }  
\newcommand{\bal}{\begin{align}}
\newcommand{\eal}{\end{align}}   
\newcommand{\bi}{\begin{itemize}}  
\newcommand{\ei}{\end{itemize}}  
\newcommand{\ben}{\begin{enumerate}}  
\newcommand{\een}{\end{enumerate}}  
\newcommand{\bc}{\begin{center}}
\newcommand{\ec}{\end{center}} 
\newcommand{\bt}{\begin{table}}
\newcommand{\et}{\end{table}}  
\newcommand{\btb}{\begin{tabular}}
\newcommand{\etb}{\end{tabular}}

\newcommand{\GeV}{\,\mathrm{GeV}}

\def\arrvline{\hfil\kern\arraycolsep\vline\kern-\arraycolsep\hfilneg}

\def\ang#1{{ \langle #1 \rangle}}

\definecolor{myblue}{HTML}{0077b6}
\definecolor{myred}{HTML}{db5a53}
\definecolor{mygreen}{HTML}{2a9d79}
\def\myr#1{{ \textcolor{myred}{#1 } }}
\def\myb#1{{ \textcolor{myblue}{#1 } }}
\def\myg#1{{ \textcolor{mygreen}{#1 } }}
\usepackage[dvipsnames]{xcolor}

\newcommand{\mzp}{m_{Z'}}
\newcommand{\neff}{N_{\rm eff}}

\allowdisplaybreaks

\let\OLDthebibliography\thebibliography
\renewcommand\thebibliography[1]{
  \OLDthebibliography{#1}
  \setlength{\parskip}{0pt}
  \setlength{\itemsep}{0pt plus 0.3ex}
}

\begin{document}

\vspace{1cm}

\begin{titlepage}

\begin{flushright}
IFT-UAM/CSIC-25-163\\
 \end{flushright}
\vspace{0.2truecm}

\begin{center}
\renewcommand{\baselinestretch}{1.8}\normalsize
\boldmath
{\LARGE\textbf{
Leptogenesis and Dark Matter \\ in an Inverse Seesaw from gauged B-L breaking}}
\unboldmath
\end{center}

\vspace{0.4truecm}

\renewcommand*{\thefootnote}{\fnsymbol{footnote}}

\begin{center}
{
Enrique Fern\'andez-Mart\'inez$^1$\footnote{\href{mailto:enrique.fernandez@csic.es}{enrique.fernandez@csic.es}}\orcidlink{0000-0002-6274-4473}, 
Ana Luisa Foguel$^2$\footnote{\href{mailto:afoguel@usp.br}{ana.foguel@unibo.it}}\orcidlink{0000-0002-4130-1200},
Xabier Marcano$^{3,4}$\footnote{\href{mailto:xabier.marcano@unibo.it}{xabier.marcano@unibo.it}}\orcidlink{0000-0003-0033-0504}, 
\\
Daniel Naredo-Tuero$^{1,5}$\footnote{\href{mailto:daniel.naredo@kit.edu}{daniel.naredo@kit.edu}}\orcidlink{0000-0002-5161-5895},
Vsevolod Syvolap$^1$\footnote{\href{mailto:vsevolod.syvolap@csic.es}{vsevolod.syvolap@csic.es}}\orcidlink{0000-0001-5829-8864
}
and Kevin A. Urqu\'ia-Calder\'on$^6$\footnote{\href{mailto:kevin.urquia@nbi.ku.dk}{kevin.urquia@nbi.ku.dk}}\orcidlink{}
}

\vspace{0.7truecm}

{\footnotesize
$^1$Instituto de F\'{\i}sica Te\'orica UAM/CSIC,\\
Universidad Aut\'onoma de Madrid, Cantoblanco, 28049 Madrid, Spain
\\[.5ex]
$^2$ Departamento de F\'isica Matem\'atica, Instituto de F\'isica \\
 Universidade de S\~ao Paulo, C. P. 66.318, 05315-970 S\~ao Paulo, Brazil
 \\[.5ex]
$^3$ Dipartimento di Fisica e Astronomia, Universit\`a di Bologna, via Irnerio 46, 40126 Bologna, Italy
\\[.5ex]
$^4$ INFN, Sezione di Bologna, viale Berti Pichat 6/2, 40127 Bologna, Italy
\\[.5ex]
$^5$ Institute for Astroparticle Physics (IAP), Karlsruhe Institute of Technology (KIT), Hermann-von-Helmholtz-Platz 1, 76344 Eggenstein-Leopoldshafen, Germany
\\[.5ex]
$^ 6$ Niels Bohr Institute, University of Copenhagen, Jagtvej 155A, DK-2200, Copenhagen, Denmark
}

\vspace*{2mm}
\end{center}

\renewcommand*{\thefootnote}{\arabic{footnote}}
\setcounter{footnote}{0}

\begin{abstract}
We study a dynamical realization of the low-scale Inverse Seesaw mechanism in which the approximate $B-L$ symmetry is gauged and spontaneously broken. Anomaly cancellation requires additional chiral fermions, one of which becomes a stable dark matter candidate after symmetry breaking, while another remains massless and contributes to dark radiation. Focusing on the regime of feeble gauge interactions, we compute the dark matter relic abundance produced via the freeze-in mechanism through the $B-L$ gauge boson and identify the parameter space consistent with cosmological and laboratory constraints. We show that the same region naturally avoids thermalization of heavy neutral leptons, preserving the viability of ARS leptogenesis. The interplay between dark matter production, dark radiation constraints, and leptogenesis requirements leads to a predictive scenario where future cosmological surveys and intensity-frontier experiments such as SHiP can probe significant portions of the viable parameter space.   
\end{abstract}

\end{titlepage}

\tableofcontents

\section{Introduction}

The discovery of neutrino masses and mixings from the neutrino oscillation phenomenon demands the extension of the Standard Model (SM) of particle physics and the existence of new physics in the neutrino sector. Among all possibilities, the inclusion of right-handed neutrinos, similarly to their charged-lepton and quark counterparts, seems the simplest and most natural option. Given their unique gauge-singlet nature, a new Majorana mass term at a new energy scale, unrelated to the Higgs mechanism and the electroweak scale, would be present unless a new symmetry prevents it. If this new scale is much larger than the electroweak scale, it also suppresses the masses of the mostly-active neutrino mass eigenstates, providing a natural explanation for their extreme lightness. This high-scale Seesaw mechanism~\cite{Minkowski:1977sc,Mohapatra:1979ia,Yanagida:1979as,Gell-Mann:1979vob} also offers an explanation for the origin of the baryon asymmetry of the Universe (BAU) through the baryogenesis via leptogenesis mechanism~\cite{Fukugita:1986hr}.

Despite its appeal, the very high-energy scale of the canonical Seesaw mechanism renders its testability a daunting task. Indeed, the new heavy neutral leptons (HNLs) are far beyond the reach of collider searches and their indirect effects are also suppressed by inverse powers of their mass. Hence, it is interesting to consider lower scale realizations of the Seesaw mechanism in which the lightness of neutrino masses may also arise naturally. In this context, the lepton number symmetry $L$, or $B-L$ as its anomaly-free extension, is violated by the light neutrino masses induced by Seesaw mechanisms. Thus, low-scale seesaw variants such as the Inverse~\cite{Mohapatra:1986aw,Mohapatra:1986bd} of Linear~\cite{Akhmedov:1995ip,Malinsky:2005bi} Seesaws may provide interesting phenomenology as well as naturally small neutrino masses through an approximate $B-L$ symmetry~\cite{Branco:1988ex,Kersten:2007vk, Abada:2007ux}. Interestingly, these low-scale realizations are not only more testable but also offer possible explanations for the BAU via ARS leptogenesis~\cite{Akhmedov:1998qx,Asaka:2005pn} as well as potential dark matter (DM) candidates~\cite{Dodelson:1993je,Shi:1998km,Abada:2014zra, Kaneta:2016vkq,Abada:2017ieq,Ghiglieri:2019kbw,Ghiglieri:2020ulj,Abada:2021yot,Abada:2023mib,Abada:2025gvc}.

In this work, we explore the possibility that the $B-L$ symmetry protecting the smallness of neutrino masses at the Inverse Seesaw is gauged, like the other fundamental symmetries in the SM, and that its breaking, which induces neutrino masses and mixings, is dynamical. As discussed in~\cite{DeRomeri:2017oxa}, see also Refs.~\cite{Khalil:2010iu,Bazzocchi:2010dt,Okada:2012np,Basso:2012ti,Kajiyama:2012xg,Cai:2014hka,Ma:2014qra,Ma:2015raa,Wang:2015saa,Okada:2016gsh,Klasen:2016qux,Escudero:2016tzx,Escudero:2016ksa,Okada:2016tci,Bandyopadhyay:2017bgh,Okada:2018ktp,Abada:2021yot}, the chiral $B-L$ charge assignment required to recover the characteristic symmetry-protected pattern of the low-scale Seesaws requires the addition of new fermion fields to cancel the chiral anomaly for the $B-L$ group. Interestingly, this leads to a dark matter candidate whose dark matter number is accidentally conserved by the $B-L$ charge assignment required for anomaly cancellation. Ref.~\cite{DeRomeri:2017oxa} explored the freeze-out production of this DM candidate and its associated phenomenology. Here, motivated by the requirement of keeping the HNLs out of thermal equilibrium for successful ARS leptogenesis~\cite{Caputo:2018zky}, we instead explore the parameter space characterized by smaller couplings of freeze-in DM production. 

In the next sections, we will study whether in this regime the model is able to address the origin of neutrino masses, produce the correct DM abundance and account for the BAU via leptogenesis. The paper is organized as follows. In Section~\ref{sec:model}, we introduce the model and discuss the new fermion degrees of freedom necessary to gauge $B-L$ in an Inverse Seesaw. We also detail the scalar sector responsible for the breaking of the symmetry and how neutrino masses can be naturally small. In Section~\ref{sec:DM} we discuss the dark matter phenomenology and identify the parts of the parameter space that would lead to the correct relic abundance. In Section~\ref{sec:Neff} we investigate the contribution to the energy density in radiation ($N_{\rm eff}$) of the other new field necessary for anomaly cancellation, which remains massless, and derive the constraints that present bounds impose on the allowed parameter space. In Section~\ref{sec:leptog} we derive the additional constraints the model needs to satisfy so as to allow for successful leptogenesis. Finally, in Section~\ref{sec:bounds} we present the allowed parameter space by all previous constraints and discuss its testability at ground-based experiments, while in Section~\ref{sec:concl} we summarize our conclusions.

\section{Dynamical Inverse Seesaw from B-L Breaking}
\label{sec:model}

Low-scale seesaw models introduce new right-handed SM singlet fermions in pairs of opposite lepton number, $N_R$ and $N'_R$, and impose that $L$ (or $B-L$) is approximately conserved, only broken by a small parameter that induces small masses for active neutrinos. In the case of the Inverse Seesaw (ISS) model, this parameter is a small Majorana mass term $\mu$ for the $N'_R$ neutrinos. More explicitly, the neutrino mass matrix in the $(\nu_L^c, N^{}_R, N'_R)$ basis is given by
\begin{equation}\label{eq:Mnu_ISS}
M_{\rm ISS} = \left(\begin{array}{ccc}
0 & y_\nu v \sfrac{}{\sqrt2} & 0 \\
y_\nu^T v \sfrac{}{\sqrt2} & 0 & M_N  \\
0 & M_N^T & \mu 
\end{array}\right) ,
\end{equation}
with $y_\nu$ the neutrino Yukawa coupling, $v$ the SM Higgs vev and $M_N$ a Dirac mass term between $N_R$ and $N'^c_R \equiv N_L$, which is allowed by all symmetries. In the $\mu\ll y_\nu v \ll M_N$ limit, the mass of light neutrinos reads
\begin{equation}
m_\nu~\simeq \frac{v^2}2 y_\nu^{} M_N^{-1} \mu (M_N^T)^{-1} y^T_\nu\,,
\end{equation}
while the active-sterile mixing is still as in the canonical seesaw $\theta\simeq vy_\nu M_N^{-1}/\sqrt2$.
Thus, the lightness of $m_\nu$ is explained by the smallness of $\mu$, while keeping $M_N$ and $y_\nu$ (and thus $\theta$) in the phenomenologically accessible regime. For example, $y_\nu\sim\mathcal O(0.1)$ and $M_N\sim\,$TeV is possible for $\mu\sim\,$keV.

Although it is technically natural to consider a small $\mu$-term, since it is protected by $B-L$, it is also interesting to explore its possible dynamical explanation, potentially originated from the breaking of a gauged $B-L$ symmetry~\cite{Khalil:2010iu,Bazzocchi:2010dt,Okada:2012np,Basso:2012ti,Kajiyama:2012xg,Cai:2014hka,Ma:2014qra,Ma:2015raa,Wang:2015saa,Okada:2016gsh,Klasen:2016qux,Escudero:2016tzx,Escudero:2016ksa,Okada:2016tci,DeRomeri:2017oxa,Bandyopadhyay:2017bgh,Okada:2018ktp,Abada:2021yot}. Due to the opposite charges of the right-handed neutrino pairs, they do not contribute to the anomaly cancellation\footnote{Anomaly cancellation does not imply any constraint in the amount of pairs. In this work, we will consider 3 of them, as it is the minimal amount to generate masses for the three light neutrinos in the ISS~\cite{Abada:2014vea}. Nevertheless, our main conclusions would not change if a different number of pairs were considered.} and, consequently, new fermions charged under $B-L$ need to be introduced in order to gauge this symmetry. Here, we will consider the minimal extension introduced in Ref.~\cite{DeRomeri:2017oxa}, consisting of 3 new fermions $\chi_R$, $\chi_L$, and $\omega$, all of them SM singlets and with $B-L$ charges as given in Table~\ref{tab:particles}.

\begin{table}[t!]
\begin{center}
\begin{tabular}{|c|c|c|c|c|c|c|c|}
\hline
Particle  & $\phi_1$  & $\phi_2$ & $N_R$ & $N'_R$ & $\chi_R$ & $\chi_L$ & $\omega$ \\[.5ex]
\hline
$U(1)_{B-L}$ charge & +1 & +2 & -1 & +1 & +5 & +4 & +4 \\
\hline
\end{tabular}
\caption{New particle content of the model~\cite{DeRomeri:2017oxa}, consisting of SM singlets but charged under $U(1)_{B-L}$. $\phi_{1,2}$ are scalar fields responsible for the spontaneous symmetry breaking of $U(1)_{B-L}$; $N_R$ and $N_R'$ are right-handed fields forming the pseudo-Dirac neutrino pairs needed for the Inverse Seesaw model; $\chi_R$ and $\chi_L$ are right- and left-handed fermions forming a stable Dirac fermion, DM candidate; and $\omega$ is a left-handed field, which remains massless.  }\label{tab:particles}
\end{center}
\end{table}

Given the $B-L$ symmetry, active neutrinos as well as the three new fermions are massless, while the $N_R$ and $N_L$ form Dirac pairs. Therefore, in order to explain neutrino oscillation data, the symmetry must be broken. This is accomplished via two new complex scalars $\phi_1$ and $\phi_2$, with $B-L$ charges of $+1$ and $+2$, respectively. The former will induce a Dirac mass term for the $\chi_R$ and $\chi_L$ fields, while the latter will generate Majorana masses for the right-handed neutrinos. More specifically, the new Lagrangian in the neutrino sector is given by\footnote{Notice that a Yukawa between $\phi_1$, $\chi_R$, and $\omega$, similar to the one with $\chi_L$, is also possible. However, without loss of generality, we work in the basis where the linear combination of these fields that couples through the Yukawa is dubbed $\chi_L$ and the orthogonal one that will remain massless is $\omega$.} 
\begin{equation}\label{eq:Lneu}
-\mathcal L_\nu = \bar L y_\nu \widetilde H N^{}_R + \overline{N^c_R} M_N N'_R + \frac12\phi_2 \overline{N^c_R} y_{\scaleto{N}{5pt}} N^{}_R + \frac12\phi^*_2\overline{N'^c_R} y'_{\scaleto{N}{5pt}} N'_R + \phi_1^*\overline{\chi_L} y_\chi \chi_R + h.c.
\end{equation}
where flavour indices have been omitted. After $B-L$ and electroweak spontaneous symmetry breaking (EWSSB), and with the conventions $\phi_j=(v_j+\varphi_j+i a_j)/\sqrt{2}$ and $H^0 = (v+h+i\varphi_Z)/\sqrt{2}$ for the neutral scalar fields around their vevs, this Lagrangian generates mass terms for the neutrinos and the new fermions. In the basis $(\nu_L^c, N^{}_R, N'_R, \chi_L^c, \chi_R)$, the mass matrix reads
\begin{equation}\label{eq:Mnu}
\renewcommand{\arraystretch}{1.1}
M_\nu = \left(\begin{array}{ccc;{3pt/6pt}cc}
0 & y_\nu v \sfrac{}{\sqrt2} & 0 & 0 & 0 \\
y_\nu^T v \sfrac{}{\sqrt2} & y_{\scaleto{N}{4pt}} v_2\sfrac{}{\sqrt{2}} & M_N & 0 & 0 \\
0 & M_N^T & y_{\scaleto{N}{4pt}}' v_2\sfrac{}{\sqrt{2}} & 0 & 0 \\
\hdashline[3pt/6pt]
0 & 0 & 0 & 0 & y_\chi v_1 \sfrac{}{\sqrt2} \\
0 & 0 & 0 & y_\chi v_1 \sfrac{}{\sqrt2}  & 0
\end{array}\right).
\end{equation}
Given the very different charge assignments (see Table~\ref{tab:particles}), the model has three decoupled sectors. First, the $(\nu_L^c, N^{}_R,N'_R)$ form an ISS sector, as in Eq.~\eqref{eq:Mnu_ISS}, with HNL masses $m_{N_i} \simeq [M_N]_{ii}$ and the $\mu$-term\footnote{A small Majorana mass is also generated for the $N_R$ fields, however we can neglect it as it induces light neutrino masses at loop level.} dynamically generated by $v_2$, which can be naturally small in part of the parameter space. Second,  $\chi_L$ and $\chi_R$ form a dark fermion, with a Dirac mass $m_{\chi}= y_\chi v_1/\sqrt{2}$ and completely stable due to an accidental global $U(1)$, corresponding to DM number, remaining after the breaking of $U(1)_{B-L}$. We explore its viability as a DM candidate in the next section. Finally, the also dark $\omega$ field remains massless, and its contribution to $N_{\rm eff}$ will impose strong constraints on the parameter space, as we will discuss in Sec.~\ref{sec:Neff}.

Regarding the scalar sector, the complete potential is now 
\begin{align}
V =& \phantom{+}\frac{m_H^2}2 H^\dagger H^{} + \frac{\lambda_H}2 (H^\dagger H^{})^2 
+ \frac{m^2_1}2\phi^*_1\phi^{}_1 + \frac{m^2_2}2 \phi^*_2\phi^{}_2 + \frac{\lambda_1}2 (\phi^*_1\phi^{}_1)^2 + \frac{\lambda_2}2 (\phi^*_2 \phi^{}_2)^2 \nonumber\\
&+ \frac{\lambda_{12}}2(\phi^*_1\phi^{}_1)(\phi^*_2\phi^{}_2) + \frac{\lambda_{1H}}2 (\phi^*_1\phi^{}_1)(H^\dagger H^{}) + \frac{\lambda_{2H}}2 (\phi^*_2\phi^{}_2)(H^\dagger H^{})
-\eta\big(\phi^2_1 \phi^*_2 + \phi^{*2}_1\phi^{}_2\big)\,.
\end{align}
In order to induce an ISS in the neutrino sector, it is interesting to consider the case where $m_H^2$ and $m_1^2$ are both negative, while $m_2^2$ is positive and large~\cite{DeRomeri:2017oxa}. In such scenario, $v_2$ is only induced by  $v_1$ via the cubic interaction $\eta$
\begin{equation}
v_2\simeq \frac{\sqrt2\, \eta\,v_1^2}{m_2^2}\,,
\end{equation} 
and can be therefore small for small $\eta$, explaining the smallness of the $\mu$-term in the ISS. This is a technically natural choice, very similar to the type-II~\cite{Magg:1980ut,Mohapatra:1980yp,Lazarides:1980nt} Seesaw, since an additional global $U(1)$ symmetry is recovered when $\eta \to 0$.
In this scenario, the mass matrix for the neutral scalar sector in the $(h, \varphi_1,\varphi_2)$ basis is
\begin{equation}
M_0^2 \simeq \left(\begin{array}{ccc}
\lambda_H v^2 & \lambda_{1H}v_1v/2 & 0 \\
\lambda_{1H}v_1v/2 & \lambda_1 v_1^2 & -\sqrt2\,\eta\, v_1 \\
0 & -\sqrt2\,\eta\, v_1 & m_2^2/2
\end{array}\right)\,.
\end{equation}
Assuming that the $h-\varphi_1$ mixing $(\alpha_1)$ is small as to comply with Higgs data~\cite{ATLAS:2019nkf,CMS:2020gsy}, and since the $\varphi_1-\varphi_2$ mixing $(\alpha_2)$ is $\eta$-suppressed, the physical masses are 
\begin{equation}
m_h^2 \simeq \lambda_H v^2\,,\qquad
m_{\varphi_1}^2 \simeq \lambda_1 v_1^2\,, \qquad
m_{\varphi_2}^2 \simeq m_2^2/2\,,
\end{equation}
with mixings
\begin{equation}
\tan\alpha_1 \simeq \frac{\lambda_{1H}}{\lambda_1}\frac v{2v_1}\,,\qquad
\tan\alpha_2 \simeq 2\frac{v_2}{v_1}\,.
\end{equation}
The scalar sector also contains a physical pseudoscalar~\cite{DeRomeri:2017oxa}
\begin{equation}
    A = \frac1{\sqrt{v_1^2+4v_2^2}}\big(2v_2a_1 - v_1a_2\big)\,,
\end{equation}
which in our scenario is also heavy, $m_A\simeq m_{\varphi_2}$, and will not impact our analysis.

Moreover, after symmetry breaking the $Z'$ gauge boson associated to the $B-L$  becomes massive, 
\begin{equation}
m_{Z'} = g_{\scaleto{B-L}{5pt}}\sqrt{v_1^2+4v_2^2}\simeq g_{\scaleto{B-L}{5pt}} v_1\,,
\end{equation}
with $g_{\scaleto{B-L}{5pt}}$ the new gauge coupling of $U(1)_{B-L}$. This massive boson will be the mediator between the SM particles and the new dark particles $\chi$ and $\omega$, with interactions\footnote{Since the $B-L$ gauge symmetry provides direct couplings to SM fermions, an additional tree-level kinetic mixing $\epsilon$ with the hypercharge gauge boson, although allowed, would not provide new phenomenology. We therefore set $\epsilon \to 0$, noting that loop-induced kinetic mixing is of order $\mathcal{O}(g_{B-L}/16\pi^2)$ and therefore subleading with respect to the direct coupling.} given by 
\begin{equation}
    \label{eq:Zp_int}
    -\mathcal{L}_{Z'}^{\scaleto{\rm int}{4.5pt}} = g_{\scaleto{B-L}{5pt}} Z^\prime_\mu\, \bigg\{\bar{\chi}\, \gamma^\mu (4 P_L + 5 P_R) \chi + 4 \,\bar{\omega} \,\gamma^\mu P_L \, \omega+  J_{B-L}^\mu \bigg\}\,,
\end{equation}
where $J_{B-L}$ is the new gauge group vector current involving SM fermions, given by
\begin{equation}
    J_{B-L}^\mu =  \sum_{f}  q^f_{\scaleto{B-L}{5pt}} \, \bar{f} \gamma^\mu \, f\, + \sum_{ \ell = e, \mu , \tau} q^{\nu_\ell}_{\scaleto{B-L}{5pt}} \,\bar \nu_\ell \gamma^\mu P_L \nu_\ell \, ,
\end{equation}
with charge assignments $(q^q_{\scaleto{B-L}{5pt}}, q^\ell_{\scaleto{B-L}{5pt}} ,q^{\nu_\ell}_{\scaleto{B-L}{5pt}}) = (\frac{1}{3},-1,-1)$. Notice that the $B-L$ charge assignment favors the interaction of the $Z'$ with the dark sector, resulting in a darker gauge boson than other standard realizations of a gauged $U(1)_{B-L}$ symmetry. This will be relevant when recasting current constraints on $B-L$ gauge bosons in Sec.~\ref{sec:bounds}.

We conclude this section by highlighting the region of the parameter space which is viable from a phenomenological point of view. As discussed later when exploring the freeze-in DM production, we will focus on the regime of tiny gauge coupling ($g_{\scaleto{B-L}{5pt}}\sim10^{-7}$) and $m_{Z'}\sim$~few GeV, which pushes the vev to very large scales $v_1\sim 10^4~{\rm TeV}$. Together with the condition of having a small $v_2$, this implies a very heavy scalar sector that will not play a significant role in the ensuing phenomenology. Regarding the dark matter candidate $\chi$, we will focus on the regime in which it is heavier than the $Z'$ and, in particular, considering $m_\chi\sim {\rm TeV}$, its Yukawa coupling is not smaller than $y_\chi\sim 10^{-3}$, at the level of the muon Yukawa coupling. As for the HNLs present in the model, we will consider their masses in the $m_{N_i}\sim 1-100\,{\rm GeV}$ range, where they can successfully account for the baryon abundance via ARS leptogenesis, and lifetimes $\tau_{N_i} \lesssim 10^{-2}$ s in order to prevent their decay products impact on the BBN. 

\section{The Dark Matter Candidate and its Abundance}
\label{sec:DM}

As previously discussed, anomaly cancellation for the new $B - L$ symmetry requires the inclusion of new fermion fields. In particular, the Dirac fermion state, $\chi$, whose left and right-handed component charge assignments are shown in Table~\ref{tab:particles}, is stable and stands as a natural Dark Matter candidate. In this section, we compute the relic abundance of this state to test the viability of the dark matter hypothesis.

Our focus is DM production via the \textbf{freeze-in mechanism} so as to preserve the success of ARS leptogenesis. As such, DM never reaches thermal equilibrium with the SM bath; instead, it is gradually populated. For definiteness, we will assume the DM abundance, together with that of the massless $\omega$ fermion and the $Z'$ mediator as well as the HNLs, to be negligible after reheating. Furthermore, we also assume the $B-L$ symmetry to be already broken at that scale. Indeed, the temperature of $B-L$ breaking is expected to be very high given the smallness of $g_{\scaleto{B-L}{5pt}}$.

The freeze-in production of DM stops once the relevant production channels become Boltzmann-suppressed. This typically takes place when the temperature $T$ of the universe drops to the order of the DM mass, $T \sim m_\chi$. In order for this mechanism to be viable, very feeble interactions between the DM and SM sectors are required, which strongly constrain the gauge couplings. At the same time, the other dark sector particles, $\omega$ and $Z'$, would also be populated through the different processes from a negligible initial abundance. For gauge coupling larger than $g_{\scaleto{B-L}{5pt}} \gtrsim 10^{-8}$, we find that the mediator eventually thermalizes with the SM bath, which in turn leads to the thermalization of $\omega$ as well.

\begin{figure}[t!] 
\centering
\includegraphics[width=0.9\textwidth]{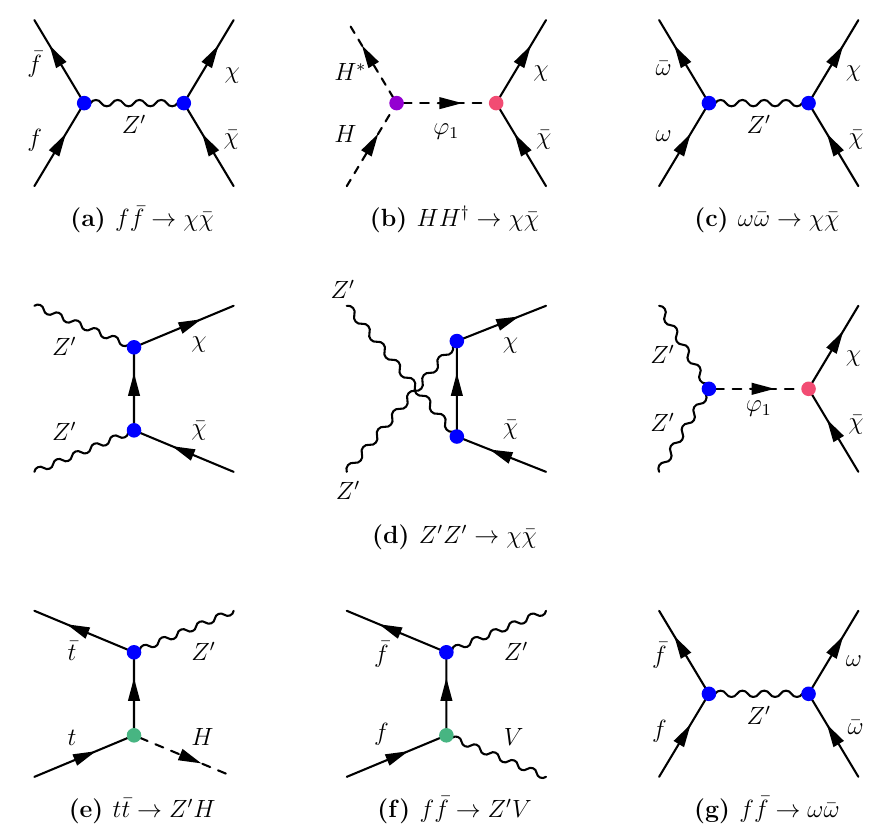}
\caption{\label{fig:BEQdiagrams} First and second rows represent the main DM production channels: s-channel for (a) SM fermion, (b) Higgs doublet, and (c) massless fermion $\omega$; and (d) $Z'$ annihilation through $t$-, $u$-, and $s$-channel processes.
Diagrams (e) and (f) represent the dominant production channels for the mediator via, respectively, top quark pair annihilation into $Z'$ and $H$, and fermion annihilation into $Z'$ and a gauge boson $V = B, W$ (before EWSSB), or $V = \gamma$ (after EWSSB). Crossed-channel contributions are not shown.
The last diagram illustrates the production or annihilation of $\omega$'s through an $s$-channel $Z'$ mediator. Vertex color code indicates coupling strengths: blue scales with $g_{\scaleto{B-L}{5pt}}$, purple with $\lambda_{1H}$, pink with the DM Yukawa coupling $y_\chi$, and green with SM couplings.
}
\end{figure}

The system of coupled Boltzmann equations that governs the evolution of the comoving number densities $Y \equiv n/s$ (with $n$ the number density and $s$ the entropy density) as a function of $x \equiv m_{\chi}/T$, for the three relevant dark sector species ($Y_{\rm DM}$, $Y_{Z'}$ and $Y_{\omega}$) is
\begin{align}
    \dv{\myb{Y_{\rm DM}}}{x} &= \frac{s}{H x} \, \frac12 (Y_{\rm DM}^{\rm eq})^2 \, \qty[  \ang{\sigma v}_{{\rm DM} \to {\rm SM}}  
    + \ang{\sigma v}_{\chi Z'}  \qty( \frac{\myg{Y_{Z'}}}{Y_{Z'}^{\rm eq}} )^2
    + \ang{\sigma v}_{\chi \omega}  \qty( \frac{\myr{Y_{ \omega}}}{Y_{ \omega}^{\rm eq}} )^2 ] \, , \label{eq:Boltzmann_YDM}\\
    \nonumber\\
    \dv{\myg{Y_{ Z' }}}{x} &= \frac{1}{H x} \, \Bigg\{\, s \, \qty( \ang{\sigma v}_{\rm top} \, (Y_{t}^{\rm eq})^2 + \ang{\sigma v}_{VZ'} \, Y_{Z'}^{\rm eq} \, Y_{V}^{\rm eq}) \qty[ 1 - \frac{\myg{Y_{Z'}}}{Y_{Z'}^{\rm eq}} ] \nonumber\\
     &\phantom{=\frac{1}{H x} \, \Bigg\{} 
     - \, \myg{Y_{Z'}}  \qty( \ang{\Gamma}_{\omega} \qty[1 - \frac{Y_{Z'}^{\rm eq}}{\myg{Y_{Z'}}} \qty( \frac{\myr{Y_{ \omega}}}{Y_{ \omega}^{\rm eq}} )^2 ] + \ang{\Gamma}_{f} \qty[1 - \frac{Y_{Z'}^{\rm eq}}{\myg{Y_{Z'}}} ]  )  \Bigg\}\, ,
      \label{eq:Boltzmann_YZ}\\
    \nonumber\\
    \dv{\myr{Y_{\omega }}}{x} &= \frac{1}{H x} \, \Bigg\{  \ang{\Gamma}_{\omega}  \myg{Y_{Z'}} \qty[1 - \frac{Y_{Z'}^{\rm eq}}{\myg{Y_{Z'}}} \qty( \frac{\myr{Y_{ \omega}}}{Y_{ \omega}^{\rm eq}} )^2 ]  - s \, \ang{\sigma v}_{\omega  f }^{\rm off} \, \myr{Y_{ \omega}}^2 \qty[1 - \qty( \frac{Y_{ \omega}^{\rm eq}}{\myr{Y_{ \omega}}} )^2 ]\, 
    \Bigg\} \, .\label{eq:Boltzmann_Yomega}
\end{align}
%
%
Notice that, since all processes involving the DM candidates are CP-conserving, $Y_{\rm DM} \equiv Y_{\chi} +Y_{\bar \chi}$ is the total DM abundance and, similarly, $Y_{\omega} = Y_{\bar \omega} $. In the above expressions, $H$ denotes the Hubble expansion rate while the equilibrium abundances are defined as $Y_i^{\rm eq} \equiv n_i^{\rm eq}/s$. The labels used for the thermally averaged cross sections times velocity, as well as for the thermally averaged decay widths, correspond to the following processes (see also Figure~\ref{fig:BEQdiagrams}):
\begin{equation}
\begin{array}{lcl@{\hspace{10pt}}|lcl}
\ang{\sigma v}_{{\rm DM} \to {\rm SM}} & : & \chi \bar \chi \to f \bar f,\; \chi \bar \chi \to H^\dagger H 
& \quad \ang{\sigma v}_{VZ'} & : & V Z' \to f \bar f  \;  \\
\ang{\sigma v}_{\chi Z'} & : & \chi \bar \chi \to Z' Z' 
& \quad \ang{\Gamma}_{\omega} & : & Z' \to \omega \bar \omega  \\
\ang{\sigma v}_{\chi \omega} & : & \chi \bar \chi \to \omega \bar \omega 
& \quad \ang{\Gamma}_{f} & : & Z' \to f \bar f   \\
\ang{\sigma v}_{\rm top} & : & t\bar t \to Z' H 
& \quad \ang{\sigma v}_{\omega f}^{\rm off} & : & \omega \bar \omega \to f \bar f\; \text{(off-shell)} \\
\end{array}
\end{equation}

For the computation of the DM abundance, we will assume a reheating temperature $T_r = 100 \,m_\chi$. We have verified that there is no significant dependence on $T_r$ except for very light mediator masses (which are disfavored by $\Delta \neff$ constraints, see Sec.~\ref{sec:Neff}). With this assumption and since the DM mass lies above the electroweak (EW) scale, the freeze-in production begins at temperatures well above the EW symmetry breaking scale. Accordingly, the gauge bosons participating in the process associated with $\ang{\sigma v}_{VZ'}$ (see diagram (f) of Fig.~\ref{fig:BEQdiagrams}) are the hypercharge boson $B$ and the $SU(2)_L$ components $W_i$.  After the temperature drops below the EW phase transition, we instead consider the process involving the photon. Regarding $\ang{\sigma v}_{\omega f}^{\rm off}$, the `off-shell' designation reflects the fact that the on-shell $Z'$ contribution is already accounted for in the decay width term. We include that decay term separately as the first contribution in Eq.~\eqref{eq:Boltzmann_Yomega}. All formulas and conventions used for the computation of thermally averaged cross sections, decay rates, equilibrium abundances\footnote{The Maxwell-Boltzmann approximation was shown to be accurate for DM production via freeze-in in Ref.~\cite{Blennow:2013jba}.}, and cosmological quantities follow those presented in Appendix B of Ref.~\cite{Foguel:2025hio}. Furthermore, given how the freeze-in mechanism is sensitive to effects from the entire thermal history, we utilized in-medium corrections. In particular, we implemented thermal corrections\footnote{The overall effect of accounting for thermal masses is $\mathcal{O}(10\%)$ on 
the final yields; however, their inclusion is important to determine which 
channels are kinematically open or closed (e.g.\ the coalescence 
$f\bar{f}\to Z'$ is forbidden at high $T$ due to fermion thermal masses) and to regulate IR divergences in $t$/$u$-channel processes.} to the SM particle masses, as it was done in \cite{Heeba:2019jho, Bringmann:2021sth}.

The dominant DM production channels are fermion annihilation, which scales as $g_{B-L}^4$, and Higgs doublet annihilation, which scales as $\lambda_{1H}^2 y_\chi^2$ and is therefore less relevant for small $\lambda_{1H}$. In addition, the annihilation of two $Z'$ bosons into dark matter can become highly relevant given that the production rate of the mediator is large. In particular, the contribution from the $Z'$ longitudinal modes dominates in $\ang{\sigma v}_{\chi Z'}$ at high $T$ and can substantially contribute to DM production even when the $Z'$ is not in thermal equilibrium. This production process was sometimes overlooked in the literature and can substantially change the predicted parameter space compatible with the observed DM abundance, as pointed out in Ref.~\cite{Eijima:2022dec, Seto:2024lik}. On the other hand, the term arising from $\omega$ annihilation is typically subdominant, since its abundance increases more slowly.

Regarding Eq.~\eqref{eq:Boltzmann_YZ} for the $Z'$ mediator, the first line contains the production terms, in particular contributions from top–antitop annihilation into $Z'H$, enhanced compared to other fermions given the larger coupling to the Higgs, and from fermion–antifermion annihilation into a gauge boson plus a $Z'$ dark mediator. Other naively expected production mechanisms, such as the direct coalescence, are in fact kinematically forbidden due to the large thermal masses acquired by fermions at high temperatures. The second line in the $Z'$ equation corresponds to the decay terms, which are responsible for the depletion of the mediator abundance after its thermalization. Here we assumed HNLs are too heavy for the $Z'$ to decay into them. Indeed, if this decay channel is open, it would tend to thermalize the HNL population potentially spoiling the leptogenesis mechanism, as will be discussed in Sec.~\ref{sec:leptog}.

Finally, in Eq.~\eqref{eq:Boltzmann_Yomega} for the massless fermion $\omega$, the only relevant production channel is the annihilation of fermions into $\omega$'s mediated by the $Z'$. Due to the very narrow width of the $Z'$, we treat its decay contribution separately from the annihilation term. This process is also responsible for thermalizing $\omega$ with the plasma, followed by its chemical decoupling once the interaction rate becomes inefficient compared to the Hubble expansion.

\begin{figure}[t!]
\begin{center}
\includegraphics[width=0.8\textwidth]{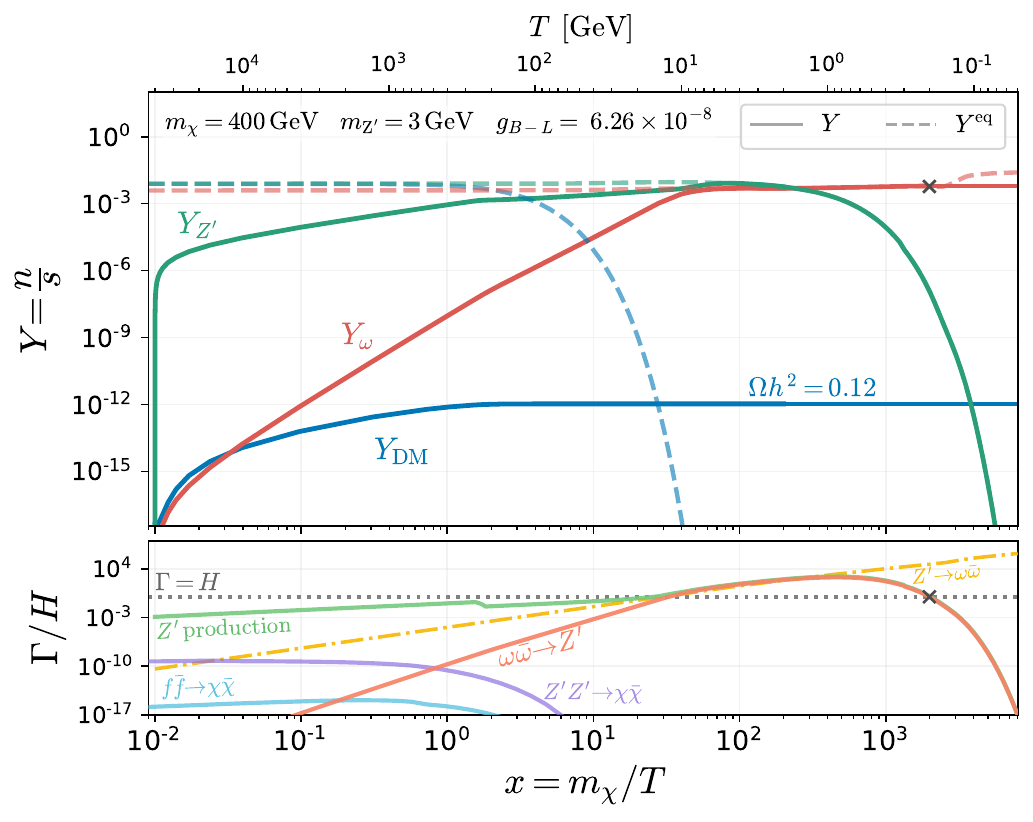}
\end{center}
\vglue -0.3 cm
\caption{\label{fig:FIcurve} The \textbf{upper panel} shows the yields of the dark sector species: $\myb{Y_{\rm DM}}$ (solid blue), $\myr{Y_{\omega}}$ (solid red), and $\myg{Y_{Z'}}$ (solid green), as a function of $x$ (lower x-axis) and temperature $T$ (upper x-axis). The masses were fixed to  $m_\chi = 400 \GeV$ and $m_{Z'} = 3 \GeV$, and the gauge coupling is chosen to reproduce the observed DM relic abundance. Dashed lines represent the corresponding equilibrium yields. The \textbf{lower panel} presents the rates, normalized to the Hubble rate, of the processes that govern the evolution of the yields: DM production via fermion annihilation (light blue) and  $Z' Z'$ annihilation (purple), $Z'$ production (light green), $\omega$ coalescence (orange), and $Z' \to \omega \bar \omega$ decays (dashed yellow). The black cross indicates the moment when the massless fermion $\omega$ freezes out.}
\end{figure}
All the qualitative behaviors discussed above are illustrated in Figure~\ref{fig:FIcurve}, which shows in the upper panel the evolution of the yields of dark matter $\myb{Y_{\rm DM}}$ (solid blue), the mediator $\myg{Y_{ Z' }}$ (solid green) and the massless fermion $\myr{Y_{ \omega }}$ (solid red) as functions of $x$, for fixed mass values $m_\chi = 400 \GeV$ and $m_{Z'} = 3 \GeV$. The dashed lines represent the corresponding equilibrium abundances, and the upper x-axis indicates the temperature scale. For this figure, we consider the regime where the scalar mixing parameter $\lambda_{1H}$ is small enough that the DM production via Higgs doublet annihilation becomes negligible, so the DM production is dominated by the channels proportional to $g_{\scaleto{B-L}{5pt}}$. Then, fixing $g_{\scaleto{B-L}{5pt}}= 6.26 \times 10^{-8}$ correctly reproduces the observed DM relic abundance $\Omega h^2 = 0.12$~\cite{Planck:2018vyg} for the chosen mass values. In addition, we fix $\lambda_1 = 1$, which for this benchmark point corresponds to a scalar mass $ m_{\phi_1} = \sqrt{\lambda_1} m_{Z'}/g_{\scaleto{B-L}{5pt}} \sim 5 \times 10^{7}  \GeV$.

The lower panel of Fig.~\ref{fig:FIcurve} displays the dominant channel rates normalized by the Hubble expansion. These rates drive the behavior of the dark sector species' yields, indicating the key moments of thermalization, freeze-in, and freeze-out. The rates are defined as follows
\begin{align}
   Z'\, {\rm production} &: \quad  \Gamma = \ang{\sigma v}_{VZ'} \, \frac{n_V^{\rm eq} \, n_{Z'}^{\rm eq}}{n_f^{\rm eq}} + \ang{\sigma v}_{\rm top} \, {n_t^{\rm eq}} +  \ang{\Gamma}_\omega \, \frac{n_{Z'}^{\rm eq} \, n_\omega}{(n_{\omega}^{\rm eq})^2} +  \ang{\Gamma}_f \, \frac{n_{Z'}^{\rm eq} \, n_\omega}{(n_{\omega}^{\rm eq})^2} \, ,\\[8pt]
     Z' \to  \omega \bar \omega  &: \quad  \Gamma = \ang{\Gamma}_\omega \, , \\[8pt]
     \omega \bar \omega \to Z'  &: \quad  \Gamma =  \ang{\sigma v}_{\omega \bar \omega \to Z'} \,  n_\omega =  \ang{\Gamma}_\omega  \, \frac{n_{Z'}^{\rm eq} \, n_\omega}{(n_{\omega}^{\rm eq})^2} \, ,\\[8pt]
    \text{DM production}~ 
  &\left\{
  \begin{aligned}
     \quad f \bar f \to \chi \bar \chi  &: \quad \Gamma = \ang{\sigma v}_{f \bar f \to \chi \bar \chi} \, n_{f}^{\rm eq} \, ,\\[4pt]
    \quad  Z' Z' \to \chi \bar \chi  &: \quad \Gamma = \ang{\sigma v}_{\chi Z'} \, n_{Z'} \, \qty(\frac{n_\chi^{\rm eq} }{n_{Z'}^{\rm eq}} )^2 \, ,
  \end{aligned}
  \right. 
\end{align}
and are labeled accordingly in the figure. When $x \sim 1$, the DM production rates (light blue and purple) begin to drop due to Boltzmann suppression, and the yield subsequently stabilizes to its final freeze-in value. As for the $Z'$ mediator, its yield exhibits a steep initial rise due to its large production rate (light green). The point at which this rate crosses the $\Gamma = H$ line marks the thermalization of the dark mediator. Since the mediator predominantly decays into $\omega$ fermions (dashed yellow), its thermalization also drives that of $\omega$. After that, the dark mediator continues to decay, following its equilibrium yield, while the rate of inverse decays of $\omega$ (orange) remains efficient. Eventually, this rate falls below Hubble expansion, indicating the chemical decoupling of the massless fermion from the plasma, or equivalently, the $\omega$ freeze-out temperature. This moment is highlighted in both panels with a black cross marker.

\begin{figure}[t!]
\begin{center}
\includegraphics[width=0.49\textwidth]{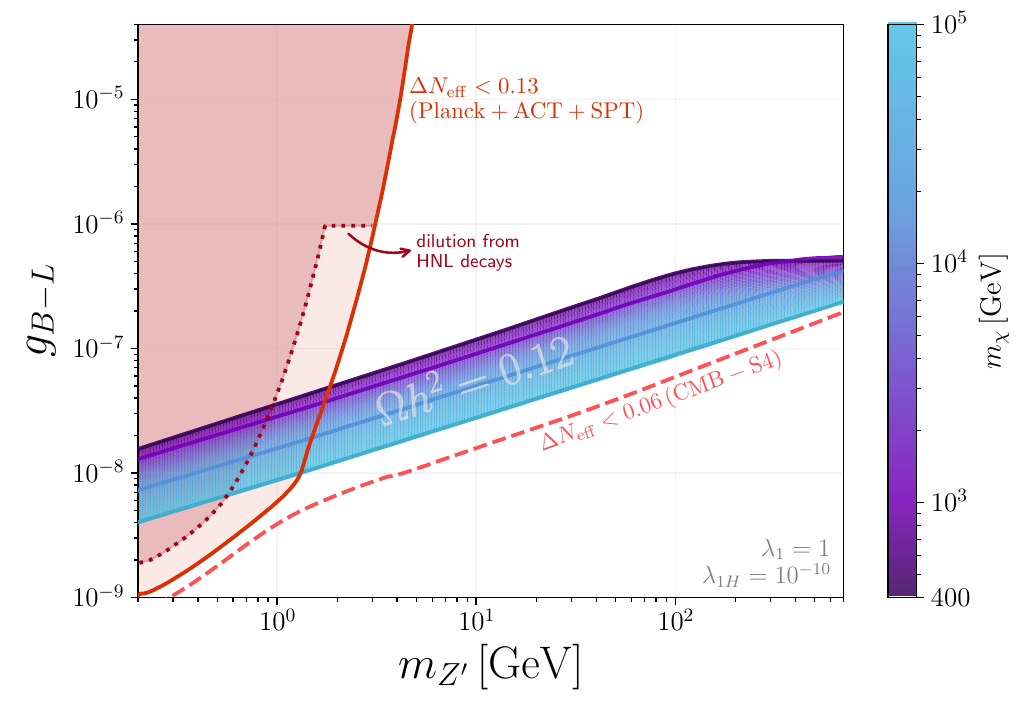}
\includegraphics[width=0.49\textwidth]{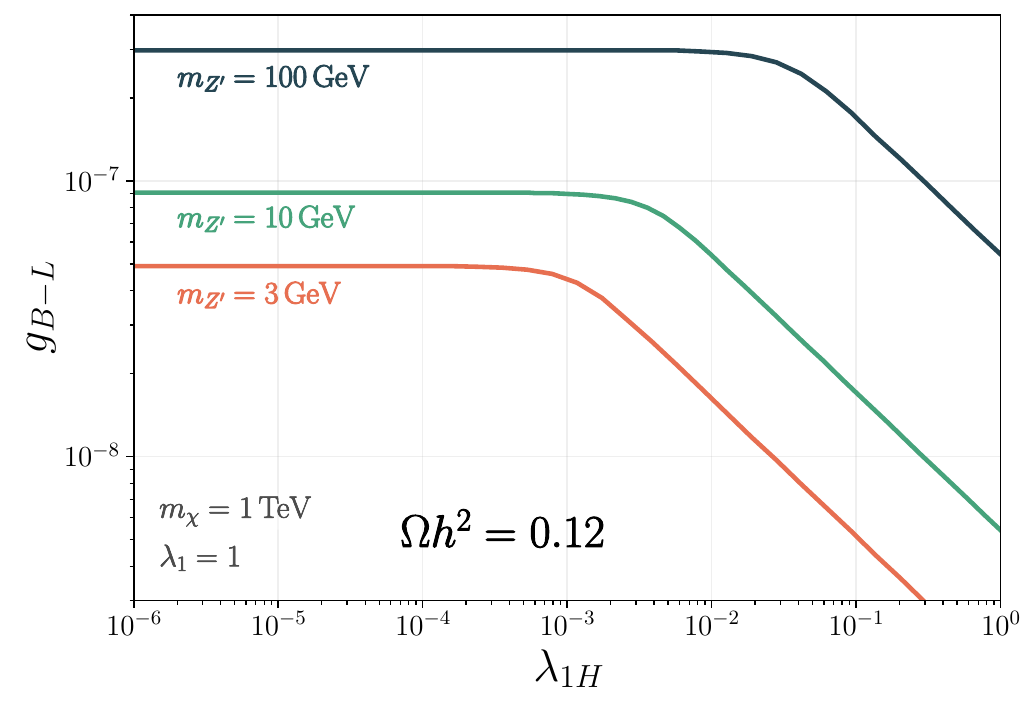}
\end{center}
\vglue -0.3 cm
\caption{\label{fig:mesa} 
Left: region of the $g_{\scaleto{B-L}{5pt}}-m_{Z'}$ plane reproducing the observed DM relic abundance for DM masses in the $(400, 10^5)$~GeV range. The red shaded area shows current bounds from $\Delta N_{\rm eff}$, including how it could be relaxed due to the dilution from HNL decays (see Sec.~\ref{sec:Neff}), while the dashed line represents future prospects.
Right: gauge coupling $g_{\scaleto{B-L}{5pt}}$ as a function of the scalar mixing $\lambda_{1H}$ that reproduces the observed DM relic abundance, for three different mediator masses and fixed values of $m_\chi$ and $\lambda_1$.
}
\end{figure}
Our results for this freeze-in scenario are summarized in the left panel of Figure~\ref{fig:mesa}, which shows the region that reproduces the observed DM relic abundance $\Omega h^2 = 0.12$~\cite{Planck:2018vyg}, in the $g_{\scaleto{B-L}{5pt}} - m_{Z'}$ plane and still for a regime dominated by gauge interactions (negligible $\lambda_{1H}$). 
The color contours represent different choices of the DM mass, ranging from $m_\chi = 400~\GeV$ to $10^5~\GeV$. The highlighted lines correspond to representative values $m_\chi/\GeV = 400$ (dark purple), $10^3$ (purple), $10^4$ (blue), and $10^5$ (light blue). In addition, the figure also displays in red the current and future projected constraints from $\Delta N_{\rm eff}$, which will be discussed in Sec.~\ref{sec:Neff}.

We conclude this section by discussing the role of the scalar potential o the DM abundance. The most relevant parameter is $\lambda_{1H}$, while the others have a negligible effect. The parameter $\lambda_1 \sim m_{\phi_1}^2 g_{B-L}^2/m_{Z'}^2$ controls the dark scalar $\phi_1$ mass. Without loss of generality, we fix $\lambda_1=1$, which, for $g_{B-L} < 10^{-6}$ and $m_{Z'} > 1$~GeV, corresponds to $m_{\phi_1} > 10^6$~GeV. Similarly, $\lambda_2$ sets the mass of the dark scalar $\phi_2$, which is also decoupled and irrelevant for the DM abundance computation. The parameters $\lambda_{12}$ and $\lambda_{2H}$, controlling respectively the $\phi_1-\phi_2$ and  Higgs-$\phi_2$ mixings, similarly do not play any role for the abundance. Finally, the parameter $\eta$ gives rise to the small Majorana mass term $\mu$, responsible for the generation of the small of neutrino masses, but does not participate in the DM phenomenology.

In the right panel of Figure~\ref{fig:mesa} we show the impact of $\lambda_{1H}$ in the DM abundance computation. For large enough values of this coupling, dark matter production via Higgs doublet annihilation (diagram (b) of Figure~\ref{fig:BEQdiagrams}) will become relevant and smaller values of the gauge coupling $g_{\scaleto{B-L}{5pt}}$ will be needed to yield the observed DM relic abundance. For a lighter mediator, the gauge coupling only starts to decrease for $\lambda_{1H} \gtrsim 10^{-3}$, whereas for heavier mediators, even larger values of $\lambda_{1H}$ are required for this effect to become significant.

As shown in the left panel of Fig.~\ref{fig:mesa}, in the regime where DM is predominantly produced via gauge interactions, the gauge coupling required to reproduce the observed relic abundance lies in the range $g_{\scaleto{B-L}{5pt}} \sim 10^{-7}$–$10^{-8}$. Interestingly, this region overlaps with the projected sensitivity of future experiments, such as SHiP, provided that $m_{Z'} \sim 0.1$–$1~\GeV$ (see Sec.~\ref{sec:bounds}). For this reason, in what follows we focus our analysis on this range of $g_{\scaleto{B-L}{5pt}}$ and neglect the effect of scalar mixing. We stress, however, that smaller values of $g_{\scaleto{B-L}{5pt}}$ can also yield the correct relic density for larger values of $\lambda_{1H}$.

\section{Dark Radiation and Constraints from $\mathbf{N_{\rm eff}}$}
\label{sec:Neff}

The presence of the massless chiral fermion $\omega$ implies an additional contribution to the radiation density of the Universe determined by Cosmic Microwave Background (CMB) and Big Bang Nucleosynthesis (BBN) observations. This contribution is conventionally normalized to that of the SM neutrinos and thus parameterized by the effective number of relativistic species $N_{\rm eff}$ as
\begin{equation}
    \label{eq:Neff}
    N_{\rm eff} \equiv \frac{8}{7}\left(\frac{11}{4} \right)^{4/3} \left( \frac{\rho_\text{rad}-\rho_\gamma}{\rho_\gamma}\right)\ ,
\end{equation}
where $\rho_\mathrm{rad}$ and $\rho_\gamma$ are the total radiation and photon energy density, respectively. Any BSM particles contributing to the energy density during the BBN or CMB epochs would affect primordial element abundances (D, He-3, He-4, Li-7, etc.) or properties of the CMB power spectrum, respectively~\cite{Yeh:2022heq,Giovanetti:2024eff}. Planck observations provide a constraint $N_{\rm eff} = 2.99^{+0.34}_{-0.33}$ at the 95~\%~C.I. when combined with BAO measurements, see Ref.~\cite{Planck:2018vyg}. Recently, new data from the ACT~\cite{ACT:2025tim} and SPT~\cite{SPT-3G:2025bzu} collaborations allowed to improve the combined CMB constraint to $N_{\rm eff}= 2.81\pm 0.24\,\,(\rm 95\% \,C.I.)$. We employ the constraint on $\Delta N_{\rm eff}\equiv N_{\rm eff}-N_{\rm eff}^{\rm SM}$ obtained by following the procedure of Ref.~\cite{ACT:2025tim} and, integrating the one-tailed $N_{\rm eff}$ posterior above $N_{\rm eff}^{\rm SM}=3.043-3.044$~\cite{Cielo:2023bqp,Drewes:2024wbw,Escudero:2025kej}, we obtain
\begin{equation}
 \Delta N_{\rm eff}<0.13 \,\,(\rm 95\%\,C.I.)\hspace{0.75cm}{\rm (Planck+ACT+SPT)}.
\end{equation}

As discussed above, the main contribution to $\Delta\neff$ stems from the production of the massless fermion $\omega$, which may take place either through freeze-in or even freeze-out depending on the parameter space. Conversely, the new heavy scalar degrees of freedom responsible for the $B-L$ symmetry breaking as well as the $Z'$ gauge boson decay much before the onset of BBN and thus do not contribute to $\neff$. HNLs, on the other hand, play a more subtle role and will be discussed at the end of this section. 

The massless fermion $\omega$ is only coupled directly to the $Z'$. As such, it can be produced in $2\to2$ processes like $i\bar{i}\to \omega\omega$, where $i$ can be either a SM fermion or one of the new particles in the model, as depicted in diagrams (c) and (g) of Figure~\ref{fig:BEQdiagrams}. However, since the abundances of the latter are generally smaller than their equilibrium abundance, $\omega$-production from the SM bath is dominant. Most importantly, due to its massless nature, the $\omega$ can be produced resonantly via the $Z'$, in stark contrast with the DM, which is produced off-resonance due to its heavy mass. As discussed in the previous section, for $g_{\scaleto{B-L}{5pt}}\gtrsim 10^{-8}$, the $\omega$ will reach thermal equilibrium around $T\sim m_{Z'}$ and its abundance will be eventually frozen-out when the temperature drops sufficiently below $m_{Z'}$. In this situation, its contribution to $\Delta \neff$ is determined by its freeze-out temperature $T_{\rm \omega FO}$, which will determine the amount of entropy dilution that will take place from the subsequent annihilations within the SM plasma. On the other hand, for smaller couplings, the $\omega$ abundance will be frozen-in and its contribution to $\neff$ cannot, in principle, be derived through the equilibrium abundance along with entropy conservation arguments.

In order to estimate $\Delta\neff$ in both of these regimes, we solve the coupled Boltzmann equations for the abundances of both the $Z'$ mediator and $\omega$ (see Eqs.~(\ref{eq:Boltzmann_YZ},~\ref{eq:Boltzmann_Yomega})) up until the $\omega$ yield, $Y_\omega$, no longer evolves. Then, assuming that the $\omega$ phase space distribution has a thermal shape, it is possible to infer its temperature today and, through it, its contribution to the energy density. When normalized to that of neutrinos, the contribution to the extra relativistic degrees of freedom is
\begin{equation}
     \Delta\neff\simeq \left(\frac{11\pi^4 g^*_{\text{s},0}}{135 \zeta(3)}  \right)^{\tfrac43} \left(Y_\omega^{\infty}\right)^{\tfrac43} = 76.3 \,\left(Y_\omega^{\infty}\right)^{\tfrac43}\,,
    \label{eq:DeltaNeff_estimate}
 \end{equation}
where $Y_\omega^\infty$ is the $\omega$ yield at $T\ll m_{Z'}$ and $g^*_{\text{s},0}$ is the effective number of relativistic degrees of freedom contributing to the entropy density today. 
The above expression is exact if the massless fermion equilibrates with the SM bath and decouples instantaneously afterwards. However, if thermalization is never reached or decoupling is non-instantaneous (which can happen for $g_{B-L} < 10^{-8}$), the $\omega$ may have a non-thermal distribution that could alter the relation above. In this scenario, solving the Boltzmann equation for the phase-space distribution would be required to compute the exact contribution to $\Delta \neff$~\cite{DEramo:2024jhn,Ovchynnikov:2024xyd}. However, as shown in \cite{DEramo:2024jhn} the contribution to $N_\mathrm{eff}$ in that case is comparable or actually lower than the thermal one and thus our estimate should be conservative and robust. Moreover, since $\omega$ generally reaches thermal equilibrium in most of the parameter space compatible with correct DM relic abundance (see Fig.~\ref{fig:mesa}), Eq.~\eqref{eq:DeltaNeff_estimate} should provide a reasonable approximation.

The resulting constraints from our analysis are shown in red in the left panel of Figure~\ref{fig:mesa}. The current constraint $\Delta\neff<0.13$ (solid red) does rule out a sizable amount of parameter space for mediators with $m_{Z'} \lesssim 2$ GeV and $g_{B-L}\gtrsim 10^{-8}$. Since a Weyl fermion decoupling after $T_{\rm QCD}$ induces $\Delta\neff\gtrsim0.4$, the latest CMB data strongly disfavors a $Z'$ below the GeV scale. Notice, however, how the $\Delta\neff$ bound substantially weakens for $g_{B-L}\lesssim10^{-8}$, since the $\omega$ no longer thermalizes and its abundance (or, equivalently, its energy density) is suppressed with respect to the equilibrium one.
For mediator masses above 2~GeV and $g_{B-L}\sim 10^{-7}-10^{-8}$, the $\omega$ freezes-out at $T_{\rm \omega FO}\simeq m_{Z'}/15\gtrsim 150\,{\rm MeV}$, before the QCD transition. Therefore, the substantial entropy dilution that takes place at $T_{\rm QCD}\sim 150$ MeV renders that part of the parameter space compatible with the bounds.

Regarding the HNLs present in the model, depending on their masses and mixings, they can have significant implications in cosmology. In particular,  they can also lead to sizable contributions to $\Delta N_{\rm eff}$, spoil BBN or contribute to the DM abundance~\cite{Dolgov:2000jw,Vincent:2014rja,Sabti:2020yrt,Boyarsky:2020dzc}. In this work, we focus on HNLs that are not excluded by the aforementioned considerations and lie within the relevant window for ARS leptogenesis~\cite{Klaric:2021cpi}. Interestingly, they can still have an impact on the $\Delta N_{\rm eff}$ contribution induced by the $\omega$. In particular, HNLs in the GeV-scale and below are produced from the SM plasma via the Dodelson-Widrow mechanism~\cite{Dodelson:1993je} while still relativistic. However, if they are long-lived, they can become non-relativistic and sizably contribute to the Universe energy density, even if their abundance is much smaller than the equilibrium one~\cite{Vincent:2014rja, Boyarsky:2021yoh}. As such, upon their decay, they can inject a substantial amount of energy into the SM bath. This can reduce the energy density fraction of the $\omega$'s and therefore, their contribution to $\Delta\neff$ if its decoupling occurs before the HNL decay\footnote{In principle, this heating of the SM bath would also reduce the DM abundance. However, as we will show later, this is an order one effect for $\Delta N_\mathrm{eff}$ and would be somewhat smaller for DM, translating into only a $\sim 10\%$ impact in the required value of $g_{B-L}$ in a particular region of the parameter space. Or, equivalently, a small shift in the DM mass range depicted in the band of allowed parameter space in Fig.~\ref{fig:mesa}.}. 

In order to quantify the level at which long-lived HNL decays can alter the constraints from $\Delta\neff$ on the $g_{\scaleto{B-L}{5pt}} - m_{Z'}$ parameter space, we solve the HNL energy density evolution as in Ref.~\cite{Vincent:2014rja}. In particular, we compute the HNL energy density relative to the SM at the temperature of their decay $T_{\rm decay}$,
\begin{equation}
    \frac{\rho_{\rm HNL}(T_{\rm decay})}{\rho_{\rm SM}(T_{\rm decay})}\equiv\alpha\,.
\end{equation}

If the freeze-out temperature of the $\omega$ lies above $T_{\rm decay}$, then the contribution of $\omega$ to $\Delta\neff$ will be suppressed by the energy injection of the HNLs. In particular, we modify the $\Delta \neff$ value obtained in the absence of HNLs in the following way
\begin{equation}
    \Delta\neff^{\rm with-HNL}=\frac{1}{1+\alpha}\Delta\neff^{\rm without-HNL}\,.
\end{equation}

Following this prescription, we scan the HNL parameter space and, for each point in the $g_{\scaleto{B-L}{5pt}} - m_{Z'}$ parameter space, we choose the $m_N-\theta^2$ value that yields the highest dilution. We restrict ourselves to the region of parameter space in which $m_{Z'}<2\,m_N$. Otherwise, HNLs can be produced resonantly from $Z'$ decays, reach equilibration, and spoil the simple production and dilution scenario presented above. Furthermore, we also restrict ourselves to heavy neutrinos that decay before light neutrinos decouple ($T_{\rm decay}\gtrsim 2$~MeV), as otherwise their decay products themselves would spoil $N_{\rm eff}$. Lastly, the picture of dominant HNL production via mixing is only accurate if their production via the $Z'$ mediator is subleading. We have verified that, for the gauge couplings relevant for DM production ($g_{\scaleto{B-L}{5pt}}\lesssim10^{-7}$), HNL production via new gauge interactions is very much suppressed with respect to production via mixing. Thus, we only show the effect of HNLs for $g_{B-L}<10^{-6}$.

The weakening of the $\Delta\neff$ bounds from the decay of long-lived HNLs is shown in the left panel of Fig.~\ref{fig:mesa} by the dark red dotted line. Since all the regions of parameter space in which $T_{\rm\omega FO}>T_{\rm QCD}$ are below the current CMB bound, the relevant effect comes from HNLs that decay between the QCD phase transition and neutrino decoupling. As shown by Fig.~\ref{fig:mesa}, the HNLs can inject enough energy in the SM bath after $\omega$-decoupling to allow slightly lower values of $m_{Z'}$. Although this effect is not very large, it opens up parameter space in the phenomenologically relevant window for future experiments such as SHiP, as we will discuss in Sec.~\ref{sec:bounds}. Conversely, we find that the HNLs that yield the highest dilution are characterized by small mixing so that they are longer-lived and hence more challenging to probe at SHiP.

Regarding future prospects of probing the model via its contribution to $\Delta\neff$, it should be noted that, in the absence of dilution from HNL decay, an $\omega$ population that decouples before the QCD transition generates $\Delta\neff\simeq 0.07$. This contribution could be tested with future CMB observatories, such as the Simons Observatory~\cite{SimonsObservatory:2025wwn} or a project with similar sensitivity to the, now no longer supported, CMB-S4 survey. While the Simons Observatory expected sensitivity of $\sigma\left(\neff\right)=0.09\,(95\%\,\rm C.L.)$~\cite{Green:2019glg}, may not be enough to fully probe this scenario, a CMB-S4-like survey could exclude $\Delta\neff>0.06$, as indicated by the red dashed line in the left panel of Fig.~\ref{fig:mesa}.

\section{Impact on ARS Leptogenesis}
\label{sec:leptog}

A very appealing feature of the seesaw mechanism is its potential to explain both neutrino masses and the baryon asymmetry of the Universe (BAU) via leptogenesis~\cite{Fukugita:1986hr}. In its simplest version, leptogenesis requires very heavy neutrinos ($m_N\gtrsim 10^{8}$~GeV~\cite{Davidson:2002qv}), whose out-of-equilibrium decays produce the necessary lepton asymmetry that gets reprocessed during EWSSB by electroweak sphaleron processes. Although successful, this canonical scenario of thermal leptogenesis is hardly testable due to the very high scales involved. However, it was later shown~\cite{Akhmedov:1998qx,Asaka:2005pn} that heavy neutrinos in the $1-100$ GeV range, frozen-in from the SM thermal bath, could generate the required lepton asymmetry through their CP-violating oscillations. Interestingly, the low-scale seesaw textures, such as the Inverse Seesaw under consideration, can naturally accommodate the parameter space for a successful leptogenesis via oscillations, also dubbed ARS leptogenesis. For various characterizations of the viable parameter space for ARS leptogenesis, we refer the reader to Refs.~\cite{Hernandez:2016kel,Drewes:2017zyw,Hernandez:2022ivz}.

A crucial requirement for successful leptogenesis via oscillations is that, due to freeze-in production, the HNLs must have feeble interactions with the thermal bath so as to prevent the thermalization of at least one of them. In the scenario under study, apart from the Yukawa interactions with the Higgs and lepton doublets that are already present in the standard case, the HNLs, carrying $B-L$ charges, couple directly to the $Z'$ gauge boson. This in turn means that there are additional interactions with the SM, which could in principle spoil ARS leptogenesis by thermalizing all HNLs before $T_{\rm EW}\simeq 140$ GeV. Indeed, these constraints have been carefully studied in Ref.~\cite{Caputo:2018zky} for several scenarios, including a gauged $B-L$. However, since our particular realization of the model is somewhat different from what was studied in Ref.~\cite{Caputo:2018zky}, we dedicate this section to revisiting the aforementioned constraints from ARS leptogenesis.

\begin{figure}[t!]
    \centering
    \includegraphics[width=0.8\textwidth]{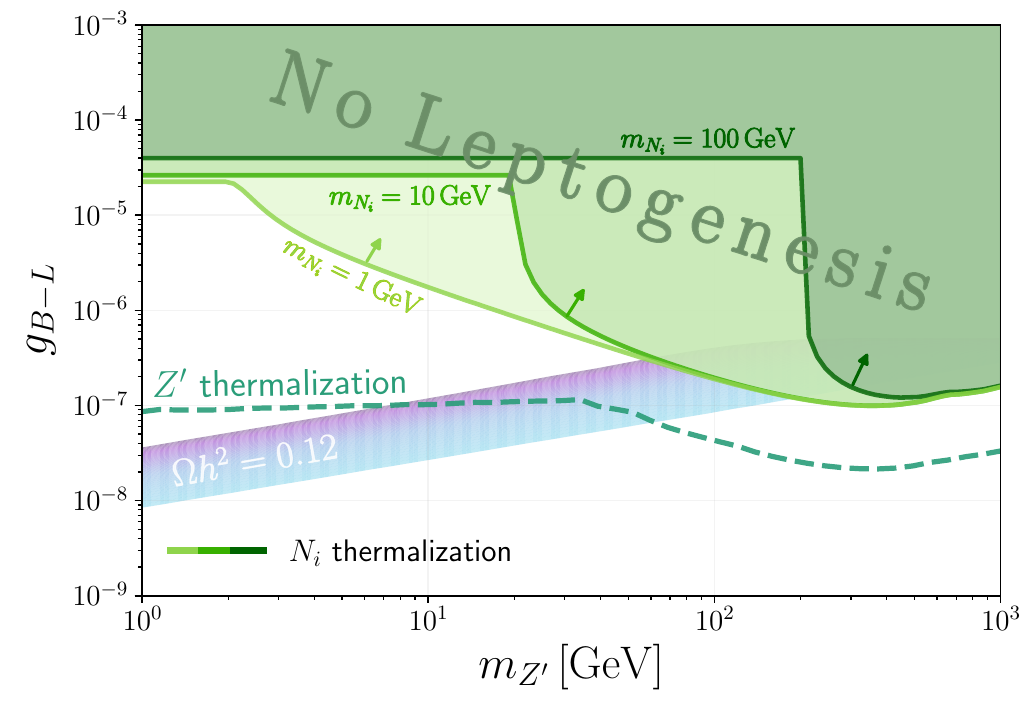}
    \caption{Parameter space where the $Z'$ mediator and HNLs thermalize with the SM plasma above or near the electroweak phase transition ($T\gtrsim 140\,{\rm GeV}$), spoiling ARS leptogenesis. The change in slope of the dashed $Z'$ line corresponds to the kinematical threshold of the coalescence channel $f\bar{f}\to Z'$, whereas the change in the solid $N_i$ lines corresponds to the threshold for the decay $Z'\to N_i \bar{N}_i$. We also show the region in which the DM abundance can be generated via freeze-in when gauge interactions dominate its production, for the same range of DM masses as in the left plot of Fig.~\ref{fig:mesa}.}
    \label{fig:leptogenesis_thermalisation}
\end{figure}

In the scenario under study, first DM production peaks and freezes-in at $T\sim m_\chi$. This has no impact on ARS leptogenesis, which typically takes place at lower temperatures. Conversely, the thermalization of the $B-L$ gauge boson plays an important role. For $T\gtrsim T_{\rm EW}$ and $m_{Z'} \lesssim 35$ GeV, the coalescence process $f\bar{f}\to Z'$ is kinematically closed due to the sizable thermal masses acquired by the SM fermions, as discussed in Sec.~\ref{sec:DM}. Instead, the main contributions to the equilibration of the $Z'$ boson are the processes depicted in diagrams (e) and (f) of Fig.~\ref{fig:BEQdiagrams}. For $m_{Z'} \gtrsim 35$ GeV, the coalescence $f\bar{f}\rightarrow Z'$ dominates the production. In Fig.~\ref{fig:leptogenesis_thermalisation} we show the region of $g_{\scaleto{B-L}{5pt}}-m_{Z'}$ parameter space (above the dashed line) in which the $B-L$ gauge boson is thermalized at $T_{\rm EW}$ or before. 
We see that for the region reproducing the correct DM abundance with a lighter mediator ($m_{Z'}\lesssim {\rm few\,GeV}$), the $Z'$ does not reach equilibrium before $T_{\rm EW}$ and thus the new interactions do not spoil ARS leptogenesis.

On the other hand, assuming that the $Z'$ is thermalized, the thermalization of HNLs depends critically on whether the decay channel $Z'\to N_i\bar{N}_i$ is kinematically allowed.
When $m_{Z'}>2m_{N_i}$, the $Z'$-decay channel dominates the production of the HNLs. In this case, the parameter space in which the HNLs are in equilibrium is given by
\begin{equation}
    \Gamma\left(Z'\to N_i \bar{N}_i\right)>H\,,
\end{equation}
where
\begin{equation}
    \Gamma\left(Z'\to N_i \bar{N}_i\right)=\dfrac{n_{Z'}^{\rm eq}}{n_{N_i}^{\rm eq}}\left\langle\Gamma_{Z'\to N_i\bar{N}_i}\right\rangle=\dfrac{n_{Z'}^{\rm eq}}{n_{N_i}^{\rm eq}}\Gamma_{Z'\to N_i\bar{N}_i}\dfrac{K_1(m_{Z'}/T)}{K_2(m_{Z'}/T)}\,,
\end{equation}
with $K_{n}$ the modified Bessel function of the second kind and $\Gamma_{Z'\to N_iN_i}$ the partial decay width
\begin{equation}
    \Gamma_{Z'\to N_i\bar{N}_i}=\dfrac{g_{\scaleto{B-L}{5pt}}^2}{12\pi\,m_{Z'}}\left(m_{Z'}^2 + 2m_{N_i}^2\right)\left(1-\frac{4m_{N_i}^2}{m_{Z'}^2}\right)^{1/2}\,.
\end{equation}

\begin{figure}[t!]
\centering
\includegraphics[width=0.9\textwidth]{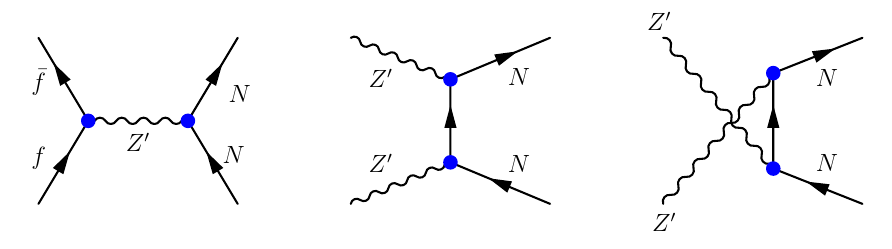}
    \caption{Main diagrams contributing to the thermalization of the heavy neutrinos $N$ when $Z'$-decays are kinematically forbidden. The $s$-channel process $Z'Z'\to\varphi_2\to \bar{N}N$ is suppressed with respect to the above diagrams by the very small $\varphi_1-\varphi_2$ mixing.}
    \label{fig:ZZ_to_NN_diagrams}
\end{figure}

Alternatively, if $Z'$-decay to HNLs is kinematically closed, the thermalization of the $N_i$ proceeds via the $ff\to N_i\bar{N}_i$ and $Z'Z'\to N_i \bar{N}_i$ processes given by the diagrams shown in Fig.~\ref{fig:ZZ_to_NN_diagrams}. A qualitative difference with respect to the scenario studied in Ref.~\cite{Caputo:2018zky} concerns the $Z'Z'\to \bar{N}N$ process, in particular the scalar-mediated $s$-channel. While in Ref.~\cite{Caputo:2018zky} the HNLs coupled directly to the scalar responsible for spontaneous $B-L$ breaking, which also provided them with a Majorana mass term, in this case the HNLs have bare Dirac mass terms, unrelated to the scale of $B-L$ breaking. Instead, they only couple directly to the $\phi_2$ scalar, which acquires a small induced vev so as to generate the Inverse Seesaw texture, as discussed in Sec.~\ref{sec:model}. Consequently, the $\varphi_2Z'Z'$ vertex is suppressed by this small vev, which means that the scalar mediated $s$-channel is suppressed by the very small $\varphi_2-\varphi_1$ mixing and therefore negligible. Moreover, in Ref.~\cite{Caputo:2018zky}, the dynamical origin of the Majorana masses of the heavy neutrinos from the spontaneous $B-L$ breaking generates a constant term in the $Z'Z'\to N_i \bar{N}_i$ cross-section for $s\gg m_{Z'},m_{N_i}$ due to the $Z'$ longitudinal modes. This constant behavior induces a strong temperature scaling for the production rate, given by $\Gamma(Z'Z'\to N_i\bar{N}_i)\propto \tfrac{g_{\scaleto{B-L}{5pt}}^4 m_{N_i}^2}{m_{Z'}^4}T^3$, which imposes tight constraints on the parameter space for a viable leptogenesis via oscillations. However,  since the HNLs are effectively Dirac in nature and their mass is not tied to the $B-L$ breaking scale, upon computing the $Z'Z'\to N_i \bar{N}_i$ diagrams, the $\propto T^3$ behavior in the production rate is absent, and it is instead roughly given by
\begin{equation}
    \Gamma\left(Z'Z'\to N_i \bar{N}_i\right)\simeq \dfrac{54\pi^3 g^4_{\scaleto{B-L}{5pt}}}{9\zeta(3)}T\log\dfrac{3T}{m_{N_i}}\,,
\end{equation}
where we have neglected $m_{Z'}$. Upon comparing this production rate with the Hubble rate
\begin{equation}
    H(T)=\sqrt{\frac{8\pi^3 g_*}{90}}\frac{T^2}{M_P}\,,
\end{equation}
where $M_P$ is the Planck mass and $g_*$ is the effective number of relativistic degrees of freedom, it is easy to see that, for $T\sim T_{\rm EW}$, this process is out of equilibrium provided that
\begin{equation}
    g_{\scaleto{B-L}{5pt}}\lesssim 10^{-4}\,.
\end{equation}

The overall picture of heavy neutrino thermalization is depicted in Fig.~\ref{fig:leptogenesis_thermalisation}, where we show as solid lines the regions of $g_{\scaleto{B-L}{5pt}} - m_{Z'}$ parameter space above which the HNLs are in equilibrium somewhere before the electroweak phase transition, for several choices of their masses. When $\mzp<2m_{N}$, the viable region for DM production is safe from spoiling the ARS leptogenesis mechanism. Interestingly, even in some regions of parameter space in which the $Z'$ would be able to decay into HNLs, our DM target is also compatible with leptogenesis via oscillations for $m_{Z'}\lesssim 50$~GeV, where the thermally averaged decay rate is suppressed for $T>T_{\rm EW}\simeq 140$ GeV due to the lightness of the $Z'$. For heavier mediators, $\mzp\gtrsim 50$ GeV it is necessary that $\mzp<2m_N$ so as to prevent thermalization via $Z'$-decays. Therefore, we conclude that overall the parameter space yielding the correct DM relic abundance remains largely compatible with successful ARS leptogenesis. In this context, ARS leptogenesis proceeds in the usual way and the values of the HNL masses and mixings for which the correct baryon abundance is obtained are reported in Refs.~\cite{Hernandez:2016kel,Drewes:2017zyw,Hernandez:2022ivz}.

\section{Current and Future Laboratory Probes}
\label{sec:bounds}

The extended particle content of the model leads to distinctive signatures at intensity-frontier experiments and, to a lesser extent, at colliders. In this section, we summarize the most relevant experimental avenues to probe the parameter space compatible with freeze-in dark matter production and ARS leptogenesis. Figures~\ref{fig:current_bounds} and \ref{fig:future_bounds} illustrate current bounds and future sensitivities, highlighting the region in which the correct relic DM abundance is obtained and including cosmological constraints from $\Delta N_{\rm eff}$ (see Sec.~\ref{sec:Neff}) and successful ARS leptogenesis (see Sec.~\ref{sec:leptog}). 

\subsection{$Z^\prime$ searches}\label{sec:Z_bounds}
\begin{figure}
    \centering
    \includegraphics[width = \linewidth]{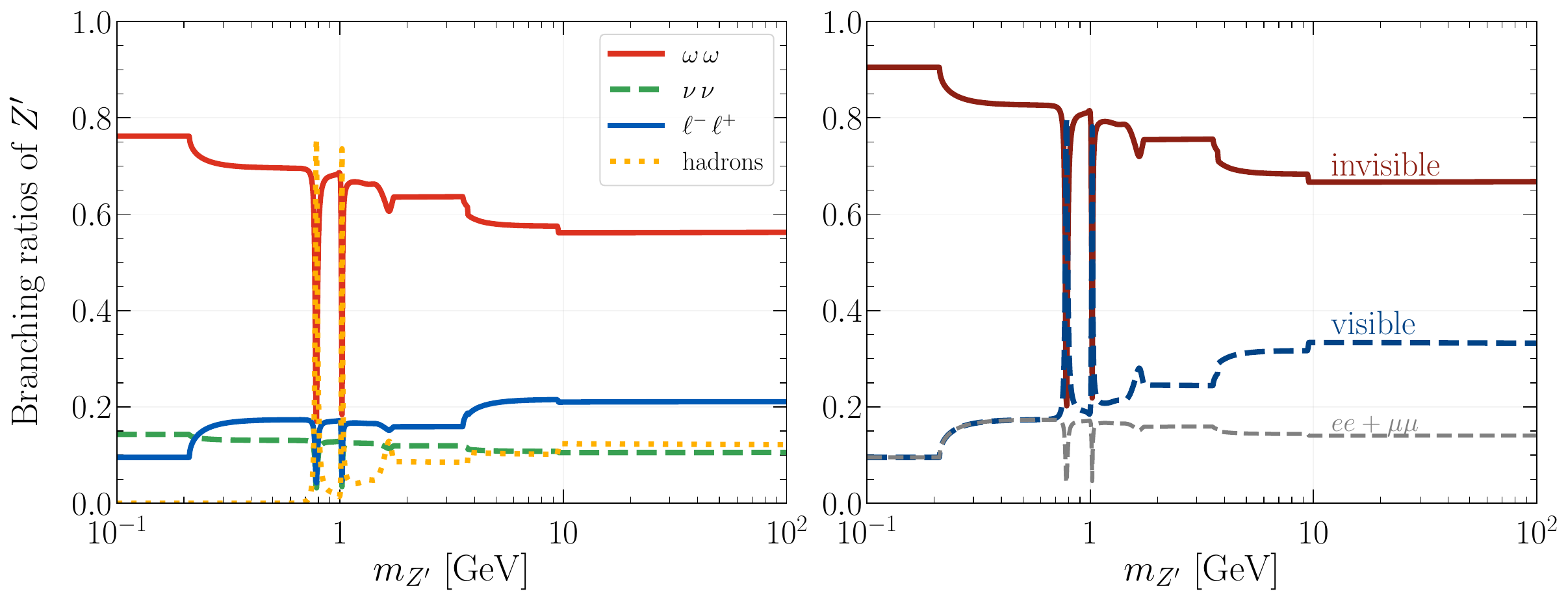}
    \caption{Branching ratios of $Z^\prime$. The panel on the left shows individual branching ratios, grouping all decays involving quark jets or QCD bound states as \textit{hadrons}. The panel on the right shows the decay products that are visible and invisible to detectors. The branching ratios were computed using a modified version of the code in \cite{Foguel:2022ppx, Foguel:2024lca}.}
    \label{fig:branching_ratio_Z}
\end{figure}

The $Z^\prime$ mediating the interactions between the SM and the dark sector can be constrained by searches for $B-L$ gauge bosons.
Nevertheless, the $Z^\prime$ in this model is characterized by having additional decays to invisible particles ($Z^\prime \to \omega\omega$), as well as larger couplings to the dark sector (see Table~\ref{tab:particles}). Consequently, between 65--90\% of the decays of the $Z^\prime$ are invisible (see Fig.~\ref{fig:branching_ratio_Z}) which makes it particularly elusive~\cite{DeRomeri:2017oxa}.
The enhanced invisible decays result in weaker bounds and sensitivities compared to other gauged $B-L$ extensions. Furthermore, this additional decay channel of the $Z^\prime$ leads to a smaller decay length, a key parameter in the context of searches for long-lived dark photons.

The different experimental searches probe complementary regions of the $g_{B-L}$ -- $m_{Z'}$ parameter space. Displaced searches probe $Z'$ with a small coupling and masses typically lower than a few GeV, with a characteristic "cigar shape", an intuition on the dependence on BSM parameters is in~\cite{Bondarenko:2019yob}. Whereas prompt searches target particles with larger coupling, but with a larger reach in mass. Beam dump experiments probe long-lived particles, whereas collider experiments can probe both displaced and prompt searches. However, we will only show prompt searches from colliders, as displaced searches from collider typically target lower masses and are thus ruled out by measurements of $\Delta N_{\mathrm{eff}}$.

\begin{figure}[t]
    \centering
    \includegraphics[width=0.9\linewidth]{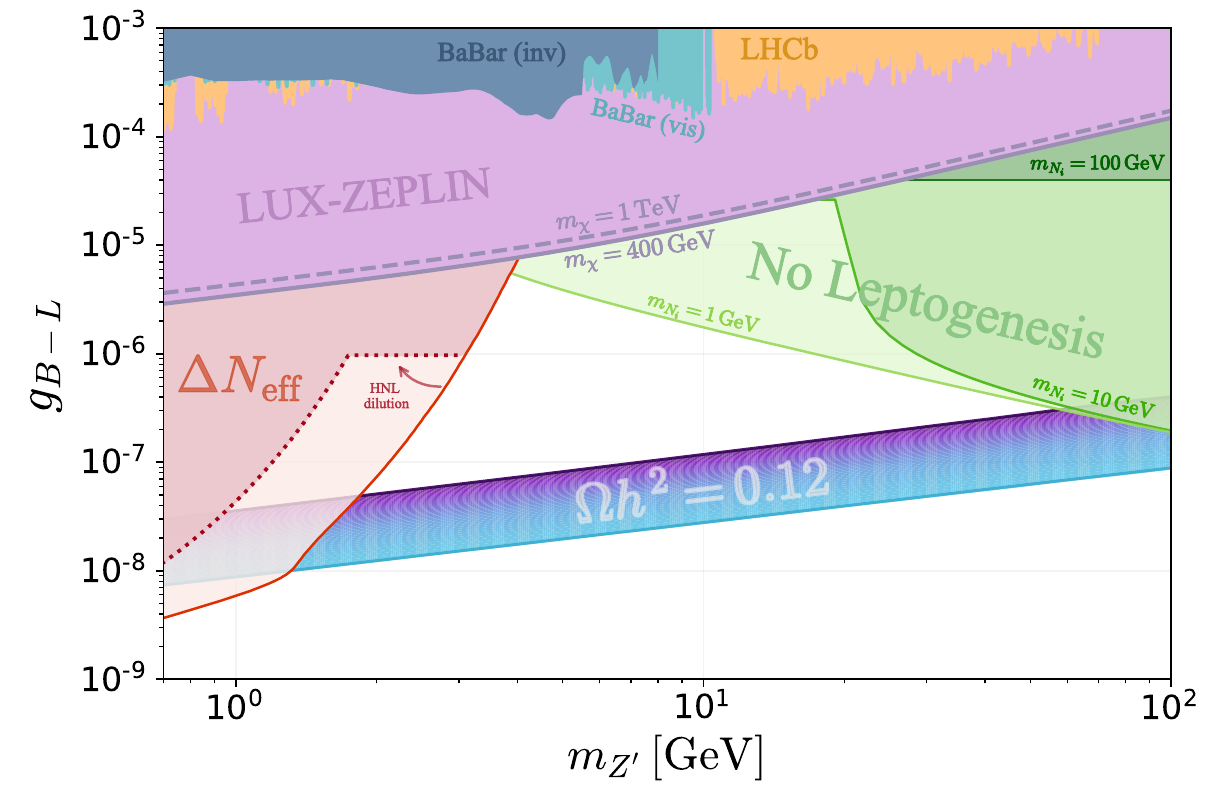}
    \caption{Current bounds on the model including the exclusion region from $\Delta N_{\mathrm{eff}}$ with and without dilution from HNLs (Sec.~\ref{sec:Neff}), as well as bounds from BaBar~\cite{BaBar:2014zli, BaBar:2017tiz}, LHCb~\cite{LHCb:2017trq}, and LUX-ZEPLIN~\cite{LZ:2018qzl} (see the main text for details on the re-casting). We also include the regions of parameter space that give the correct DM abundance for different DM masses (Sec.~\ref{sec:DM}), and that cannot provide leptogenesis (Sec.~\ref{sec:leptog}).}
    \label{fig:current_bounds}
\end{figure}

\begin{figure}[t]
    \centering
    \includegraphics[width=0.9\linewidth]{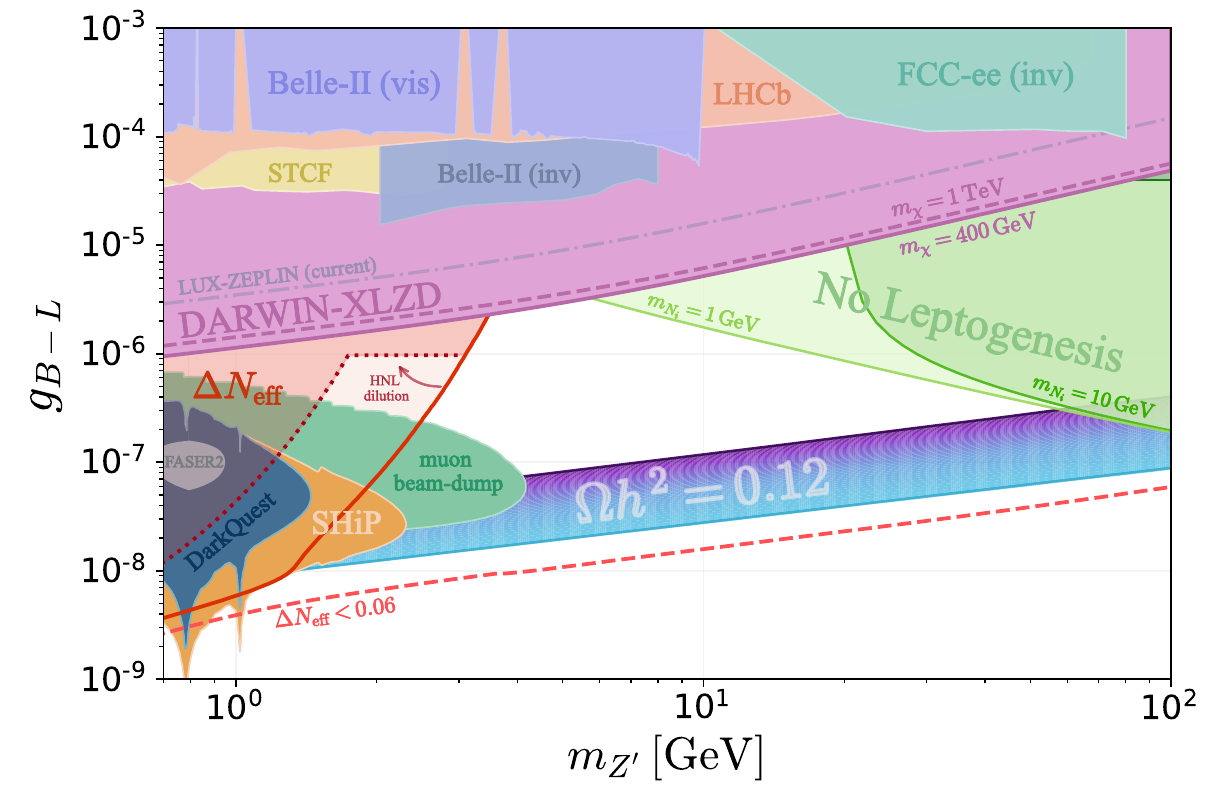}
    \caption{
    Same as Fig.~\ref{fig:current_bounds} but with prospective sensitivities from the STCF~\cite{Zhang:2019wnz}, Belle-II~\cite{Belle-II:2018jsg, Belle-II:2022cgf}, LHCb~\cite{Ilten:2015hya, Ilten:2016tkc}, FCC-ee~\cite{Airen:2024iiy}, DARWIN-XLZD~\cite{Baudis:2024jnk}, FASER2~\cite{FASER:2018eoc}, DarkQuest~\cite{Apyan:2022tsd}, SHiP~\cite{Alekhin:2015byh, SHiP:2021nfo}, and a hypothetical muon beam dump facility~\cite{Cesarotti:2022ttv, Cesarotti:2023sje} (see the main text for details on the re-casting). Also included future prospects for $\Delta N_{\mathrm{eff}}$.
    }
    \label{fig:future_bounds}
\end{figure}

The most relevant experimental probes, displayed in Fig.~\ref{fig:current_bounds} for current bounds and in Fig.~\ref{fig:future_bounds} for future sensitivities, are:
\begin{itemize}
    \item \textbf{Proton fixed-target experiments} \\
    Proton fixed-target experiments can produce a $Z^\prime$ via meson decays, proton bremsstrahlung $p\,\mathcal{N} \to p\,\mathcal{N}\,Z^\prime$ (where $\mathcal{N}$ is the target nucleus), and Drell-Yan $p\,p\to Z^\prime$ (for the treatment of the production cross-section see e.g.~\cite{Carloni:2011kk, Gorbunov:2014wqa, Foroughi-Abari:2024xlj}). The subsequent decay of these $Z'$ is then searched for at detectors located downstream of the target, providing particularly good sensitivity to very small couplings $g_{\scaleto{B-L}{5pt}} \sim 10^{-7} - 10^{-8}$ for $m_{Z^\prime}$ around \si{GeV} masses and below.

    Experiments like CHARM~\cite{CHARM:1985anb}, LSND~\cite{Batell:2009di}, Nu-CAL \cite{Blumlein:2011mv, Blumlein:2013cua} already place constraints on the model, although in a region of parameter space already constrained by its contribution to $\Delta N_{\mathrm{eff}}$. In the future, the proposed DarkQuest facility~\cite{Apyan:2022tsd}, and the now-approved SHiP experiment~\cite{Alekhin:2015byh, SHiP:2021nfo} will be capable of searching for heavier $Z^\prime$s.
    Their sensitivities were computed with the \texttt{SensCalc} code~\cite{Ovchynnikov:2023cry} and using a custom-made model with the correct decay widths and branching ratios predicted by the scenario under study.
    We conclude that the SHiP experiment will probe a maximal mass of $m_{Z^\prime} \simeq \SI{2.3}{GeV}$ at a coupling of $g_{\scaleto{B-L}{5pt}} \sim \num{2e-8}$, well within the region accepted by $\Delta N_{\mathrm{eff}}$ and that could generate the correct DM abundance. This is especially relevant if the bounds from $\Delta N_{\mathrm{eff}}$ are relaxed due to the HNL dilution explained in Sec.~\ref{sec:Neff}.

    \item \textbf{Lepton fixed-target experiments} \\
    Lepton beam dump experiments can produce a $Z^\prime$ via a bremsstrahlung process $\ell\,\mathcal{N} \to \ell\,\mathcal{N} Z^\prime$ (see \cite{Bjorken:2009mm, Liu:2017htz, Kirpichnikov:2021jev}). 
    The vast majority of lepton beam dump experiments search for visible signals, but invisible searches are also possible, where the signal is the final lepton that has lost significant energy due to the bremsstrahlung event. 

    Previous and ongoing experiments already place bounds on the $Z^\prime$ parameter space. 
    This includes experiments like E-137~\cite{Bjorken:1988as} and E-141~\cite{Riordan:1987aw}, which have searched for both long-lived and prompt hidden particles, probing couplings from $g_{\scaleto{B-L}{5pt}}\sim \num{e-3}-\num{e-8}$.
    The ongoing NA64~\cite{NA64:2016oww, NA64:2024nwj} experiment primarily searches for invisible decays, for sub-GeV $Z^\prime$s, and with couplings as low as $g_{\scaleto{B-L}{5pt}} \sim \num{e-6}$. However, 
    all of these experiments place bounds in regions of the parameter space that are ruled out by $\Delta N_{\mathrm{eff}}$. 

    An ambitious proposal for a muon beam dump experiment on a hypothetical muon collider was discussed in Refs.~\cite{Cesarotti:2022ttv, Cesarotti:2023sje}, which could search for long-lived mediators. 
    The sensitivity of this muon beam dump experiment, shown in Fig.~\ref{fig:future_bounds}, was reproduced using the WW approximation of the cross-section (see, e.g.~\cite{Kirpichnikov:2021jev}), with one of the set-ups examined in \cite{Cesarotti:2023sje}. The setup consisted of a lead target, with a beam energy of $E_\mu = \SI{1.5}{TeV}$, a target length of $L_{\mathrm{tar}} = \SI{5.0}{m}$, a shield length of $L_{\mathrm{sh}} = \SI{10.0}{m}$, a detector length of $L_{\mathrm{det}} = \SI{100.0}{m}$, and $N_\mu = \num{e22}$ muons on target. 
    We see that it could probe up to a mass of $m_{Z^\prime} \simeq \SI{4}{GeV}$ for a coupling of $g_{\scaleto{B-L}{5pt}} \simeq \num{e-7}$, covering a new part of the region reproducing the correct DM abundance.

    \item \textbf{Hadron and lepton colliders} \\
    Colliders may produce the $Z^\prime$ via bremsstrahlung, resonantly via a Drell-Yan process, or in heavy meson decays. Its decays can generate an observable signal, such as a dilepton or dihadron final state. 
    Invisible searches are also possible via initial state radiation of photons or jets.

    The parameter space is already constrained by searches at the $e^+e^-$ colliders BaBar~\cite{BaBar:2014zli} and LEP~\cite{Fox:2011fx}, as well as by the LHC detectors LHCb~\cite{LHCb:2017trq} and CMS~\cite{CMS:2019buh}; 
    and will be further explored by the next Belle-II runs~\cite{Belle-II:2018jsg, Belle-II:2022cgf, Ferber:2022ewf, Jaeckel:2023huy}, and LHCb runs~\cite{Ilten:2015hya, Ilten:2016tkc}.
    The proposed low-energy $e^+e^-$ collider, the Super-Tau Charm Factory (STCF) experiment~\cite{Zhang:2019wnz}, 
    as well as higher energy ones like the FCC-ee, and muon collider~\cite{Airen:2024iiy} will have significant sensitivities to $Z^\prime$s. 
    Unfortunately, collider experiments can only probe up to $g_{\scaleto{B-L}{5pt}} > \num{e-5}$, not reaching the feeble couplings needed in our freeze-in regime.
    Nevertheless, forward physics detectors associated with colliders, such as FASER, which is \SI{480}{m} away from the ATLAS detector at LHC, is better suited to search for long-lived particles \cite{FASER:2018eoc}. The FASER setup, however, can only probe sub-GeV masses, which are already ruled out by $\Delta N_{\mathrm{eff}}$. The bounds and sensitivities in Figs.~\ref{fig:current_bounds} and \ref{fig:future_bounds} were recasted with the same methods presented in~\cite{Ilten:2018crw}, which were recently validated in~\cite{Foguel:2024lca}. The only exception was the FASER sensitivity, which was obtained with a modified version of the \texttt{FORESEE} code~\cite{Kling:2021fwx}, which uses the correct widths and branching ratios for the model under study.
\end{itemize}

Altogether, beam dump experiments provide the best opportunity to reach regions of the parameter space that comply with cosmological bounds and that predict the correct relic DM abundance, given their suitability for searches for feebly interacting particles. However, as shown by Fig.~\ref{fig:future_bounds}, there are also parts of the parameter space in which the correct DM and baryon abundances are obtained, whose testability in terrestrial facilities is very challenging. Indeed, for $m_{Z^\prime}\gtrsim 4{\,\rm GeV}$, the $Z'$ is both too heavy for intensity experiments and too feebly interacting for colliders. In this window, the main observable signature will be in the form of a shift on $\Delta N_{\rm eff}$, in reach of future cosmological surveys. 

\subsection{Direct DM detection}
Interactions between the DM particle, $\chi$, and the rest of the SM can either be mediated by the scalar $\varphi_1$ or by the $Z^\prime$ gauge boson. 
In the region of parameter space we have focused on, $\varphi_1$ is decoupled, and interactions between the SM and $\chi$ are dominated by $Z^\prime$ mediated channels. 

These interactions can mediate spin-independent (SI) scatterings between $\chi$ and nuclei, $\cal N$. 
Current DM experiments that are sensitive to heavy DM already place bounds on the parameter space of the model; the most stringent ones come from LUX-ZEPLIN (LZ) \cite{LZ:2018qzl}. 
Different planned experiments will have a significantly larger range of sensitivity, in particular, the proposed
Darwin/XLZD \cite{Baudis:2024jnk} will probe the smallest couplings. 

The interaction terms for direct DM detection are shown in Eq.~\eqref{eq:Zp_int}.
From here we can derive the low-energy operator, where the only relevant terms are the vector-vector interactions between $\chi$ and quarks, 
\begin{align}
    \mathcal{O}^{\mathrm{SI}}_{\scaleto{\chi \mathcal{N}}{6pt}} = \frac{3g_{\scaleto{B-L}{5pt}}^2}{2 m_{Z^\prime}^2} \left[ \bar{\chi} \gamma^\mu \chi \right]\,\left[\bar{q} \gamma_\mu q\right],
\end{align}
as the axial terms are velocity suppressed (see e.g.~\cite{Fitzpatrick:2012ix, Anand:2013yka, DelNobile:2021wmp}).
The resulting total scattering cross-section between a nucleus with $Z$ protons, and $A-Z$ neutrons, written in terms of the reduced mass of the $\chi-\mathcal{N}$ system $\mu_{\scaleto{\chi \mathcal{N}}{6pt}} = \frac{m_\chi m_{\scaleto{\mathcal{N}}{3pt}}}{m_\chi + m_{\scaleto{\mathcal{N}}{3pt}}}$ and maximum transferred momentum $q_{\mathrm{max}} = 2 \mu_{\scaleto{\chi \mathcal{N}}{6pt}} v$, where $v$ is the relative velocity between DM and the target, is (see, e.g.~\cite{Anand:2013yka, DelNobile:2021wmp})
\begin{equation}
    \label{eq:}
    \sigma^{\mathrm{SI}}_{\scaleto{\chi \mathcal{N}}{6pt}} = \bar{\sigma}^{\mathrm{SI}}_{\scaleto{\chi \mathcal{N}}{6pt}}\,A^2 =  \frac{81}{4} \frac{\mu_{\scaleto{\chi \mathcal{N}}{6pt}}^2}{\pi} \frac{g_{\scaleto{B-L}{5pt}}^4}{m_{Z^\prime}^2\left(q_{\mathrm{max}}^2 + m_{Z^\prime}^2 \right)} A^2\,,
\end{equation}
where $\bar{\sigma}^{\mathrm{SI}}_{\chi \mathcal{N}}$ is the cross section per nucleon. The transferred momentum, $q_{\mathrm{max}}$ is in the order of hundreds of $\si{MeV}$ for Xenon, and tens of $\si{MeV}$ for Argon, for a DM velocity of $v = \SI{232}{\kilo\meter\per\second}$.

For Xenon targets, and assuming a DM mass $m_\chi \simeq \SI{400}{GeV}$, and with $g_{\scaleto{B-L}{5pt}} = \num{5e-8}$ and $m_{Z^\prime} = \SI{2}{GeV}$, the cross section is of order $\bar{\sigma}_{\chi-\mathrm{Xe}}^{\mathrm{SI}}\sim\mathcal{O}(\SI{e-54}{cm^{-2}})$, six orders of magnitude below the expected sensitivity of DARWIN/XLZD \cite{Baudis:2024jnk}.
For larger values of $m_{Z^\prime}$, the expected cross section will become even smaller, whereas for larger values of $m_\chi$, the sensitivity of the different experiments decreases. Therefore, the hypothetical observation of a signal by DARWIN/XLZD, or any other direct DM search experiment would point to a different DM candidate than the one discussed here. Similarly, a signal is not expected through indirect searches for the potential products of DM annihilation. Indeed, in this DM mass window, indirect searches are less constraining than their direct counterparts~\cite{Arrenberg:2013rzp}. This is also the case of self-interactions mediated by the $Z'$, which are too suppressed to be observable~\cite{Tulin:2017ara,Adhikari:2022sbh} 

\subsection{HNL searches}

Our model predicts the existence of HNLs, which we consider in the range of $m_N\sim(1-100)$~GeV in order to explain the BAU via ARS leptogenesis.
Search strategies for such HNLs based on their mixings to active neutrinos have been extensively studied, with current bounds and prospective sensitivities in the $\theta^2 - m_N$ plane comprehensively reviewed in e.g.~\cite{Atre:2009rg,Bolton:2019pcu,Fernandez-Martinez:2023phj}.

With HNLs charged under the new gauge group, the new interactions mediated by the $Z'$ induce new production and decay channels. Moreover, the preference of the $Z'$ to decay invisibly into the $\omega$ (see Fig.~\ref{fig:branching_ratio_Z}) could potentially hinder the HNL from usual searches.
Nevertheless, the extremely small couplings around $10^{-7}$ required for the freeze-in DM production imply that the new channels are always negligible, and thus the HNL phenomenology is dominated by their mixing to active neutrinos. Consequently, the HNLs of our model can be tested as in usual ARS scenarios leading to the correct lepton asymmetry~\cite{Hernandez:2016kel,Hernandez:2022ivz}, where experiments such as SHiP~\cite{Alekhin:2015byh} and FCC-ee~\cite{Blondel:2014bra} are particularly interesting.

Nevertheless, our model does show an interesting interplay between the new dark sector and the HNLs. As discussed in Sec.~\ref{sec:Neff}, HNL late decays could relax the $\Delta N_{\mathrm{eff}}$ bounds, widening the region with correct relic DM abundance that SHiP could probe. Therefore, it will be interesting to explore a simultaneous and correlated observation of both $Z'$ and HNLs at SHiP. 
Unfortunately, relaxing $\Delta N_{\mathrm{eff}}$ requires the HNLs to be particularly long-lived, with small mixings with active neutrinos, and therefore more challenging to detect at SHiP.

\section{Conclusions}
\label{sec:concl}

Given the complexity of the SM particle content, it is interesting to consider the phenomenology of extended dark sectors with non-minimal structure or interactions. In this context, three portals of SM gauge singlets have been identified that may connect, at the lowest possible order, the visible and dark sectors: the Higgs portal~\cite{Patt:2006fw}, the photon and $Z$ portal~\cite{Holdom:1985ag} and the neutrino portal~\cite{Falkowski:2009yz,Lindner:2010rr,GonzalezMacias:2015rxl,Blennow:2019fhy}. The phenomenology of these separate options has been extensively studied, but, as advocated by~\cite{Ballett:2019pyw,Ballett:2019cqp}, it is more natural to expect that several or all of them will be present in a given motivated SM extension. The SM extension studied here is an example of the latter.  

Indeed, when gauging the $B-L$ symmetry that stabilizes the light neutrino masses in an Inverse Seesaw mechanism, anomaly cancellation requires the existence of a dark sector connected to the SM via the $Z'$ associated with $B-L$, as well as through the scalar sector~\cite{DeRomeri:2017oxa}. This option connects the solution to the origin and smallness of neutrino masses to the nature and origin of the DM component of the Universe. Motivated by the possibility of also explaining the baryon asymmetry of the universe via the ARS leptogenesis mechanism, we have explored here the part of parameter space characterized by very small $B-L$ gauge coupling, so it avoids the equilibration of the Inverse Seesaw HNLs that could spoil leptogenesis. In this regime, DM production would occur via freeze-in.

We find that the parameter space where the correct DM relic abundance is produced via freeze-in is characterized by values of the $B-L$ coupling $g_{\scaleto{B-L}{5pt}} \lesssim 10^{-7}$ and $m_{Z'} \gtrsim 2$~GeV. Indeed, for lighter $Z'$, the new dark massless fermion $\omega$ decouples too late, below the QCD phase transition, and contributes too much to $\neff$. This constraint can be slightly relaxed for some regions of the parameter space where the HNL late decays may dilute the frozen $\omega$ energy density with respect to the SM thermal bath, which would increase the parameter space of the $B-L$ gauge boson potentially testable at SHiP. 

Larger $g_{\scaleto{B-L}{5pt}}$ couplings are allowed for heavier $Z'$. This is in contrast to some previous results where the required value of $g_{\scaleto{B-L}{5pt}}$ was constant in $m_{Z'}$. This is indeed the case when only the s-channel exchange of the $Z'$ is considered, but, in agreement with Ref.~\cite{Eijima:2022dec, Seto:2024lik}, we find that the inclusion of the $Z' Z' \leftrightarrow \chi \bar \chi$ channel changes this behaviour. Smaller $B-L$ couplings could also lead to the correct DM relic abundance if complemented by production through the scalar sector. Unfortunately, at low energies, the scalar sector is decoupled and this second option is not directly testable. 

All in all, the testability of this freeze-in model at beam-dumps, collider or direct DM searches is challenging, but beam dump experiments such as SHiP will be able to probe still unconstrained regions of the viable parameter space. Moreover, the present constraint in $\neff$ already excludes a significant part of the parameter space and improving these measurements can potentially rule out the entire region where DM is dominantly produced by the $Z'$ mediator. 

A complementary probe is through the usual HNL searches at beam dump facilities, which are already probing the part of the HNL parameter space interesting for ARS leptogenesis. This study was restricted to the part of the parameter space where the $Z'$ cannot decay to HNLs. If this channel is open, ARS leptogenesis and the HNL contribution to $\neff$ could be significantly altered and is thus left for future exploration.

To summarize, if the approximate $B-L$ symmetry that explains the smallness of neutrino masses in the inverse Seesaw mechanism is gauged, a DM candidate naturally appears due to anomaly cancellation. We find that, in the regime of very small $B-L$ coupling, the correct relic abundance can be obtained via freeze-in and BAU via ARS leptogenesis is unaffected. Although this scenario is challenging to fully probe through laboratory experiments, improving our determination of $\neff$ would offer the most sensitive means of exploring it.


\paragraph{Acknowledgments.} 
KAUC wants to thank Oleg Ruchayskiy, Jean Loup-Tastet, and especially Inar Timiryasov for helpful discussions.
This project has received support from the European Union’s Horizon 2020 research and innovation programme under the Marie Skłodowska-Curie grant agreement No 101086085 - ASYMMETRY, and from the Spanish Research Agency (Agencia Estatal de Investigaci\'on) through Grant IFT Centro de Excelencia Severo Ochoa No CEX2020-001007-S and grant PID2022-137127NB-I00 funded by MCIN/AEI/10.13039/501100011033 and by “European Union NextGenerationEU/PRTR''.  
The work of XM is funded by the Italian Ministry of Universities and Research (MUR) and the European Union -- Next Generation EU, Missione~4 Componente 1 CUP J33C24003210006 - NEWTRINOS and by the Italian INFN program on Theoretical Astroparticle Physics (TAsP). ALF is supported by Funda\c{c}\~ao de Amparo \`a Pesquisa do Estado de S\~ao Paulo (FAPESP) under the contracts 2022/04263-5, and 2024/06544-7. ALF and KAUC thank the Institute of Theoretical Physics (IFT-UAM) in Madrid for the warm hospitality at the beginning of this work. The work of DNT was supported by the Spanish MIU through the National Program FPU (grant number FPU20/05333) and by the Alexander von Humboldt Foundation. VS acknowledges support from the Spanish Research Agency through grant CNS2023-145338 funded by MCIN/AEI/10.13039/501100011033 and by “European Union NextGenerationEU/PRTR''. This work is partially funded by the European Commission – NextGenerationEU, through Momentum CSIC Programme: Develop Your Digital Talent. We acknowledge HPC support by Emilio Ambite, staff hired under the Generation D initiative, promoted by Red.es, an organisation attached to the Spanish Ministry for Digital Transformation and the Civil Service, for the attraction and retention of talent through grants and training contracts, financed by the Recovery, Transformation and Resilience Plan through the EU’s Next Generation funds.

\bibliographystyle{JHEP} 
\bibliography{biblio}

@article{Foguel:2025hio,
    author = "Foguel, Ana Luisa and Funchal, Renata Zukanovich and Frigerio, Michele",
    title = "{Vector dark matter with non-abelian kinetic mixing}",
    eprint = "2510.26765",
    archivePrefix = "arXiv",
    primaryClass = "hep-ph",
    month = "10",
    year = "2025"
}

@article{Fukugita:1986hr,
    author = "Fukugita, M. and Yanagida, T.",
    title = "{Baryogenesis Without Grand Unification}",
    reportNumber = "RIFP-641",
    doi = "10.1016/0370-2693(86)91126-3",
    journal = "Phys. Lett. B",
    volume = "174",
    pages = "45--47",
    year = "1986"
}

@article{Boyarsky:2021yoh,
    author = "Boyarsky, Alexey and Ovchynnikov, Maksym and Sabti, Nashwan and Syvolap, Vsevolod",
    title = "{When feebly interacting massive particles decay into neutrinos: The Neff story}",
    eprint = "2103.09831",
    archivePrefix = "arXiv",
    primaryClass = "hep-ph",
    reportNumber = "KCL-2021-05",
    doi = "10.1103/PhysRevD.104.035006",
    journal = "Phys. Rev. D",
    volume = "104",
    number = "3",
    pages = "035006",
    year = "2021"
}

@article{Branco:1988ex,
    author = "Branco, G. C. and Grimus, W. and Lavoura, L.",
    title = "{The Seesaw Mechanism in the Presence of a Conserved Lepton Number}",
    reportNumber = "IFM-11/88",
    doi = "10.1016/0550-3213(89)90304-0",
    journal = "Nucl. Phys. B",
    volume = "312",
    pages = "492--508",
    year = "1989"
}

@article{Drewes:2024wbw,
    author = "Drewes, Marco and Georis, Yannis and Klasen, Michael and Wiggering, Luca Paolo and Wong, Yvonne Y. Y.",
    title = "{Towards a precision calculation of N $_{eff}$ in the Standard Model. Part III. Improved estimate of NLO contributions to the collision integral}",
    eprint = "2402.18481",
    archivePrefix = "arXiv",
    primaryClass = "hep-ph",
    reportNumber = "CPPC-2024-01, MS-TP-24-06",
    doi = "10.1088/1475-7516/2024/06/032",
    journal = "JCAP",
    volume = "06",
    pages = "032",
    year = "2024"
}

@article{Kersten:2007vk,
	Archiveprefix = {arXiv},
	Author = {Kersten, J{\"o}rn and Smirnov, Alexei {\relax Yu}.},
	Doi = {10.1103/PhysRevD.76.073005},
	Eprint = {0705.3221},
	Journal = {Phys. Rev.},
	Pages = {073005},
	Primaryclass = {hep-ph},
	Slaccitation = {%%CITATION = ARXIV:0705.3221;%%},
	Title = {{Right-Handed Neutrinos at CERN LHC and the Mechanism of Neutrino Mass Generation}},
	Volume = {D76},
	Year = {2007}}

@article{Abada:2007ux,
	Archiveprefix = {arXiv},
	Author = {Abada, A. and Biggio, C. and Bonnet, F. and Gavela, M. B. and Hambye, T.},
	Doi = {10.1088/1126-6708/2007/12/061},
	Eprint = {0707.4058},
	Journal = {JHEP},
	Pages = {061},
	Primaryclass = {hep-ph},
	Reportnumber = {FTUAM-07-12, IFT-UAM-CSIC-07-41, LPT-ORSAY-07-34, ULB-TH-07-27},
	Slaccitation = {%%CITATION = ARXIV:0707.4058;%%},
	Title = {{Low energy effects of neutrino masses}},
	Volume = {12},
	Year = {2007}}

@article{Mohapatra:1986aw,
	title        = {{Mechanism for Understanding Small Neutrino Mass in Superstring Theories}},
	author       = {Mohapatra, R. N.},
	year         = 1986,
	journal      = {Phys. Rev. Lett.},
	volume       = 56,
	pages        = {561--563},
	doi          = {10.1103/PhysRevLett.56.561},
	slaccitation = {%%CITATION = PRLTA,56,561;%%}
}

@article{Mohapatra:1986bd,
    author = "Mohapatra, R. N. and Valle, J. W. F.",
    title = "{Neutrino Mass and Baryon Number Nonconservation in Superstring Models}",
    reportNumber = "MdDP-PP-86-127",
    doi = "10.1103/PhysRevD.34.1642",
    journal = "Phys. Rev. D",
    volume = "34",
    pages = "1642",
    year = "1986"
}

@article{Akhmedov:1995ip,
	title        = {{Left-right symmetry breaking in NJL approach}},
	author       = {Akhmedov, Evgeny K. and Lindner, Manfred and Schnapka, Erhard and Valle, J. W. F.},
	year         = 1996,
	journal      = {Phys. Lett. B},
	volume       = 368,
	pages        = {270--280},
	doi          = {10.1016/0370-2693(95)01504-3},
	eprint       = {hep-ph/9507275},
	archiveprefix = {arXiv},
	reportnumber = {IC-95-125, TUM-HEP-221-95, MPI-PHT-95-35, FTUV-95-34, IFIC-95-36}
}

@article{Dodelson:1993je,
    author = "Dodelson, Scott and Widrow, Lawrence M.",
    title = "{Sterile-neutrinos as dark matter}",
    eprint = "hep-ph/9303287",
    archivePrefix = "arXiv",
    reportNumber = "FERMILAB-PUB-93-057-A",
    doi = "10.1103/PhysRevLett.72.17",
    journal = "Phys. Rev. Lett.",
    volume = "72",
    pages = "17--20",
    year = "1994"
}

@article{Shi:1998km,
    author = "Shi, Xiang-Dong and Fuller, George M.",
    title = "{A New dark matter candidate: Nonthermal sterile neutrinos}",
    eprint = "astro-ph/9810076",
    archivePrefix = "arXiv",
    doi = "10.1103/PhysRevLett.82.2832",
    journal = "Phys. Rev. Lett.",
    volume = "82",
    pages = "2832--2835",
    year = "1999"
}

@article{Abada:2014zra,
    author = "Abada, Asmaa and Arcadi, Giorgio and Lucente, Michele",
    title = "{Dark Matter in the minimal Inverse Seesaw mechanism}",
    eprint = "1406.6556",
    archivePrefix = "arXiv",
    primaryClass = "hep-ph",
    reportNumber = "LPT-ORSAY-14-32, SISSA-35-2014-FISI",
    doi = "10.1088/1475-7516/2014/10/001",
    journal = "JCAP",
    volume = "10",
    pages = "001",
    year = "2014"
}

@article{Abada:2017ieq,
    author = "Abada, Asmaa and Arcadi, Giorgio and Domcke, Valerie and Lucente, Michele",
    title = "{Neutrino masses, leptogenesis and dark matter from small lepton number violation?}",
    eprint = "1709.00415",
    archivePrefix = "arXiv",
    primaryClass = "hep-ph",
    reportNumber = "DESY-17-124, CP3-17-23, LPT-Orsay-17-35, LPT-ORSAY-17-35",
    doi = "10.1088/1475-7516/2017/12/024",
    journal = "JCAP",
    volume = "12",
    pages = "024",
    year = "2017"
}

@article{Ghiglieri:2019kbw,
    author = "Ghiglieri, J. and Laine, M.",
    title = "{Sterile neutrino dark matter via GeV-scale leptogenesis?}",
    eprint = "1905.08814",
    archivePrefix = "arXiv",
    primaryClass = "hep-ph",
    reportNumber = "CERN-TH-2019-062",
    doi = "10.1007/JHEP07(2019)078",
    journal = "JHEP",
    volume = "07",
    pages = "078",
    year = "2019"
}

@article{Ghiglieri:2020ulj,
    author = "Ghiglieri, J. and Laine, M.",
    title = "{Sterile neutrino dark matter via coinciding resonances}",
    eprint = "2004.10766",
    archivePrefix = "arXiv",
    primaryClass = "hep-ph",
    doi = "10.1088/1475-7516/2020/07/012",
    journal = "JCAP",
    volume = "07",
    pages = "012",
    year = "2020"
}

@article{Abada:2021yot,
    author = "Abada, Asmaa and Bernal, Nicol\'as and Hern\'andez, Antonio E. C\'arcamo and Marcano, Xabier and Piazza, Gioacchino",
    title = "{Gauged inverse seesaw from dark matter}",
    eprint = "2107.02803",
    archivePrefix = "arXiv",
    primaryClass = "hep-ph",
    reportNumber = "TUM-HEP 1351/21, PI/UAN-2021-692FT",
    doi = "10.1140/epjc/s10052-021-09535-5",
    journal = "Eur. Phys. J. C",
    volume = "81",
    number = "8",
    pages = "758",
    year = "2021"
}

@article{Abada:2023mib,
    author = "Abada, A. and Arcadi, G. and Lucente, M. and Piazza, G. and Rosauro-Alcaraz, S.",
    title = "{Thermal effects in freeze-in neutrino dark mater production}",
    eprint = "2308.01341",
    archivePrefix = "arXiv",
    primaryClass = "hep-ph",
    doi = "10.1007/JHEP11(2023)180",
    journal = "JHEP",
    volume = "11",
    pages = "180",
    year = "2023"
}

@article{Abada:2025gvc,
    author = "Abada, A. and Arcadi, G. and Lucente, M. and Rosauro-Alcaraz, S.",
    title = "{Testable dark matter solution within the seesaw mechanism}",
    eprint = "2503.20017",
    archivePrefix = "arXiv",
    primaryClass = "hep-ph",
    month = "3",
    year = "2025"
}

@article{DeRomeri:2017oxa,
    author = "De Romeri, Valentina and Fernandez-Martinez, Enrique and Gehrlein, Julia and Machado, Pedro A. N. and Niro, Viviana",
    title = "{Dark Matter and the elusive $Z^\prime$ in a dynamical Inverse Seesaw scenario}",
    eprint = "1707.08606",
    archivePrefix = "arXiv",
    primaryClass = "hep-ph",
    reportNumber = "FERMILAB-PUB-17-285-T, FTUAM-17-14, IFT-UAM-CSIC-17-070",
    doi = "10.1007/JHEP10(2017)169",
    journal = "JHEP",
    volume = "10",
    pages = "169",
    year = "2017"
}

@article{Magg:1980ut,
    author = "Magg, M. and Wetterich, C.",
    title = "{Neutrino Mass Problem and Gauge Hierarchy}",
    reportNumber = "CERN-TH-2829",
    doi = "10.1016/0370-2693(80)90825-4",
    journal = "Phys. Lett. B",
    volume = "94",
    pages = "61--64",
    year = "1980"
}

@article{Mohapatra:1980yp,
    author = "Mohapatra, Rabindra N. and Senjanovic, Goran",
    title = "{Neutrino Masses and Mixings in Gauge Models with Spontaneous Parity Violation}",
    reportNumber = "FERMILAB-PUB-80-061-THY, FERMILAB-PUB-80-061-T",
    doi = "10.1103/PhysRevD.23.165",
    journal = "Phys. Rev. D",
    volume = "23",
    pages = "165",
    year = "1981"
}

@article{ATLAS:2019nkf,
    author = "Aad, Georges and others",
    collaboration = "ATLAS",
    title = "{Combined measurements of Higgs boson production and decay using up to $80$ fb$^{-1}$ of proton-proton collision data at $\sqrt{s}=$ 13 TeV collected with the ATLAS experiment}",
    eprint = "1909.02845",
    archivePrefix = "arXiv",
    primaryClass = "hep-ex",
    reportNumber = "CERN-EP-2019-097",
    doi = "10.1103/PhysRevD.101.012002",
    journal = "Phys. Rev. D",
    volume = "101",
    number = "1",
    pages = "012002",
    year = "2020"
}

@article{Bolton:2019pcu,
    author = "Bolton, Patrick D. and Deppisch, Frank F. and Bhupal Dev, P. S.",
    title = "{Neutrinoless double beta decay versus other probes of heavy sterile neutrinos}",
    eprint = "1912.03058",
    archivePrefix = "arXiv",
    primaryClass = "hep-ph",
    doi = "10.1007/JHEP03(2020)170",
    journal = "JHEP",
    volume = "03",
    pages = "170",
    year = "2020"
}

@article{Hernandez:2016kel,
    author = "Hern{\'a}ndez, P. and Kekic, M. and L{\'o}pez-Pav{\'o}n, J. and Racker, J. and Salvado, J.",
    title = "{Testable Baryogenesis in Seesaw Models}",
    eprint = "1606.06719",
    archivePrefix = "arXiv",
    primaryClass = "hep-ph",
    doi = "10.1007/JHEP08(2016)157",
    journal = "JHEP",
    volume = "08",
    pages = "157",
    year = "2016"
}

@article{Vincent:2014rja,
    author = "Vincent, Aaron C. and Martinez, Enrique Fernandez and Hern{\'a}ndez, Pilar and Lattanzi, Massimiliano and Mena, Olga",
    title = "{Revisiting cosmological bounds on sterile neutrinos}",
    eprint = "1408.1956",
    archivePrefix = "arXiv",
    primaryClass = "astro-ph.CO",
    reportNumber = "IFIC-14-53, FTUAM-14-32, IFT-UAM-CSIC-14-075",
    doi = "10.1088/1475-7516/2015/04/006",
    journal = "JCAP",
    volume = "04",
    pages = "006",
    year = "2015"
}

@article{Hernandez:2022ivz,
    author = "Hernandez, Pilar and Lopez-Pavon, Jacobo and Rius, Nuria and Sandner, Stefan",
    title = "{Bounds on right-handed neutrino parameters from observable leptogenesis}",
    eprint = "2207.01651",
    archivePrefix = "arXiv",
    primaryClass = "hep-ph",
    reportNumber = "IFIC/22-20, FTUV-22-0704.1758",
    doi = "10.1007/JHEP12(2022)012",
    journal = "JHEP",
    volume = "12",
    pages = "012",
    year = "2022"
}

@article{Atre:2009rg,
    author = "Atre, Anupama and Han, Tao and Pascoli, Silvia and Zhang, Bin",
    title = "{The Search for Heavy Majorana Neutrinos}",
    eprint = "0901.3589",
    archivePrefix = "arXiv",
    primaryClass = "hep-ph",
    reportNumber = "FERMILAB-PUB-08-086-T, NSF-KITP-08-54, MADPH-06-1466, DCPT-07-198, IPPP-07-99",
    doi = "10.1088/1126-6708/2009/05/030",
    journal = "JHEP",
    volume = "05",
    pages = "030",
    year = "2009"
}

@article{CMS:2020gsy,
    collaboration = "CMS",
    title = "{Combined Higgs boson production and decay measurements with up to 137 fb$^{-1}$ of proton-proton collision data at $\sqrt s$ = 13 TeV}",
    reportNumber = "CMS-PAS-HIG-19-005",
    year = "2020"
}

@article{Patt:2006fw,
    author = "Patt, Brian and Wilczek, Frank",
    title = "{Higgs-field portal into hidden sectors}",
    eprint = "hep-ph/0605188",
    archivePrefix = "arXiv",
    reportNumber = "MIT-CTP-3745",
    month = "5",
    year = "2006"
}

@article{Holdom:1985ag,
    author = "Holdom, Bob",
    title = "{Two U(1)'s and Epsilon Charge Shifts}",
    reportNumber = "UTPT-85-30",
    doi = "10.1016/0370-2693(86)91377-8",
    journal = "Phys. Lett. B",
    volume = "166",
    pages = "196--198",
    year = "1986"
}

@article{Ballett:2019cqp,
    author = "Ballett, Peter and Hostert, Matheus and Pascoli, Silvia",
    title = "{Neutrino Masses from a Dark Neutrino Sector below the Electroweak Scale}",
    eprint = "1903.07590",
    archivePrefix = "arXiv",
    primaryClass = "hep-ph",
    reportNumber = "IPPP/19/21",
    doi = "10.1103/PhysRevD.99.091701",
    journal = "Phys. Rev. D",
    volume = "99",
    number = "9",
    pages = "091701",
    year = "2019"
}

@article{Blennow:2019fhy,
    author = "Blennow, M. and Fernandez-Martinez, E. and Olivares-Del Campo, A. and Pascoli, S. and Rosauro-Alcaraz, S. and Titov, A. V.",
    title = "{Neutrino Portals to Dark Matter}",
    eprint = "1903.00006",
    archivePrefix = "arXiv",
    primaryClass = "hep-ph",
    reportNumber = "FTUAM-19-5, IFT-UAM/CSIC-19-19, IPPP/19/17",
    doi = "10.1140/epjc/s10052-019-7060-5",
    journal = "Eur. Phys. J. C",
    volume = "79",
    number = "7",
    pages = "555",
    year = "2019"
}

@article{Lindner:2010rr,
    author = "Lindner, Manfred and Merle, Alexander and Niro, Viviana",
    title = "{Enhancing Dark Matter Annihilation into Neutrinos}",
    eprint = "1005.3116",
    archivePrefix = "arXiv",
    primaryClass = "hep-ph",
    doi = "10.1103/PhysRevD.82.123529",
    journal = "Phys. Rev. D",
    volume = "82",
    pages = "123529",
    year = "2010"
}

@article{GonzalezMacias:2015rxl,
    author = "Gonzalez Macias, Vannia and Wudka, Jose",
    title = "{Effective theories for Dark Matter interactions and the neutrino portal paradigm}",
    eprint = "1506.03825",
    archivePrefix = "arXiv",
    primaryClass = "hep-ph",
    doi = "10.1007/JHEP07(2015)161",
    journal = "JHEP",
    volume = "07",
    pages = "161",
    year = "2015"
}

@article{Ballett:2019pyw,
    author = "Ballett, Peter and Hostert, Matheus and Pascoli, Silvia",
    title = "{Dark Neutrinos and a Three Portal Connection to the Standard Model}",
    eprint = "1903.07589",
    archivePrefix = "arXiv",
    primaryClass = "hep-ph",
    reportNumber = "IPPP/19/19/FTPI-MINN-20-17, IPPP/19/19",
    doi = "10.1103/PhysRevD.101.115025",
    journal = "Phys. Rev. D",
    volume = "101",
    number = "11",
    pages = "115025",
    year = "2020"
}

@article{Falkowski:2009yz,
    author = "Falkowski, Adam and Juknevich, Jose and Shelton, Jessie",
    title = "{Dark Matter Through the Neutrino Portal}",
    eprint = "0908.1790",
    archivePrefix = "arXiv",
    primaryClass = "hep-ph",
    reportNumber = "RU-NHETC-09-15",
    month = "8",
    year = "2009"
}

@article{Escudero:2016ksa,
    author = "Escudero, Miguel and Rius, Nuria and Sanz, Ver{\'o}nica",
    title = "{Sterile Neutrino portal to Dark Matter II: Exact Dark symmetry}",
    eprint = "1607.02373",
    archivePrefix = "arXiv",
    primaryClass = "hep-ph",
    doi = "10.1140/epjc/s10052-017-4963-x",
    journal = "Eur. Phys. J. C",
    volume = "77",
    number = "6",
    pages = "397",
    year = "2017"
}

@article{Escudero:2016tzx,
    author = "Escudero, Miguel and Rius, Nuria and Sanz, Ver{\'o}nica",
    title = "{Sterile neutrino portal to Dark Matter I: The $U(1)_{B-L}$ case}",
    eprint = "1606.01258",
    archivePrefix = "arXiv",
    primaryClass = "hep-ph",
    reportNumber = "FTUV-16-0419, IFIC-16-32",
    doi = "10.1007/JHEP02(2017)045",
    journal = "JHEP",
    volume = "02",
    pages = "045",
    year = "2017"
}

@article{Ma:2015raa,
    author = "Ma, Ernest and Srivastava, Rahul",
    title = "{Dirac or inverse seesaw neutrino masses from gauged $B–L$ symmetry}",
    eprint = "1504.00111",
    archivePrefix = "arXiv",
    primaryClass = "hep-ph",
    doi = "10.1142/S0217732315300207",
    journal = "Mod. Phys. Lett. A",
    volume = "30",
    number = "26",
    pages = "1530020",
    year = "2015"
}

@article{Ma:2014qra,
    author = "Ma, Ernest and Srivastava, Rahul",
    title = "{Dirac or inverse seesaw neutrino masses with $B-L$ gauge symmetry and $S_3$ flavor symmetry}",
    eprint = "1411.5042",
    archivePrefix = "arXiv",
    primaryClass = "hep-ph",
    doi = "10.1016/j.physletb.2014.12.049",
    journal = "Phys. Lett. B",
    volume = "741",
    pages = "217--222",
    year = "2015"
}

@article{Basso:2012ti,
    author = "Basso, Lorenzo and Fischer, Oliver and van der Bij, J. J.",
    title = "{Natural Z' model with an inverse seesaw mechanism and leptonic dark matter}",
    eprint = "1207.3250",
    archivePrefix = "arXiv",
    primaryClass = "hep-ph",
    reportNumber = "FR-PHENO-2012-015",
    doi = "10.1103/PhysRevD.87.035015",
    journal = "Phys. Rev. D",
    volume = "87",
    number = "3",
    pages = "035015",
    year = "2013"
}

@article{Khalil:2010iu,
    author = "Khalil, Shaaban",
    title = "{TeV-scale gauged B-L symmetry with inverse seesaw mechanism}",
    eprint = "1004.0013",
    archivePrefix = "arXiv",
    primaryClass = "hep-ph",
    doi = "10.1103/PhysRevD.82.077702",
    journal = "Phys. Rev. D",
    volume = "82",
    pages = "077702",
    year = "2010"
}

@article{Bazzocchi:2010dt,
    author = "Bazzocchi, Federica",
    title = "{Minimal Dynamical Inverse See Saw}",
    eprint = "1011.6299",
    archivePrefix = "arXiv",
    primaryClass = "hep-ph",
    doi = "10.1103/PhysRevD.83.093009",
    journal = "Phys. Rev. D",
    volume = "83",
    pages = "093009",
    year = "2011"
}

@article{Cai:2014hka,
    author = "Cai, Yi and Chao, Wei",
    title = "{The Higgs Seesaw Induced Neutrino Masses and Dark Matter}",
    eprint = "1408.6064",
    archivePrefix = "arXiv",
    primaryClass = "hep-ph",
    reportNumber = "ACFI-T14-15",
    doi = "10.1016/j.physletb.2015.08.026",
    journal = "Phys. Lett. B",
    volume = "749",
    pages = "458--463",
    year = "2015"
}

@article{Bandyopadhyay:2017bgh,
    author = "Bandyopadhyay, Priyotosh and Chun, Eung Jin and Mandal, Rusa",
    title = "{Implications of right-handed neutrinos in $B-L$ extended standard model with scalar dark matter}",
    eprint = "1707.00874",
    archivePrefix = "arXiv",
    primaryClass = "hep-ph",
    reportNumber = "IITH-PH-0001-17, IMSC-2017-07-05",
    doi = "10.1103/PhysRevD.97.015001",
    journal = "Phys. Rev. D",
    volume = "97",
    number = "1",
    pages = "015001",
    year = "2018"
}

@article{Okada:2016tci,
    author = "Okada, Nobuchika and Okada, Satomi",
    title = "{$Z^\prime$-portal right-handed neutrino dark matter in the minimal U(1)$_X$ extended Standard Model}",
    eprint = "1611.02672",
    archivePrefix = "arXiv",
    primaryClass = "hep-ph",
    reportNumber = "YGHP16-06",
    doi = "10.1103/PhysRevD.95.035025",
    journal = "Phys. Rev. D",
    volume = "95",
    number = "3",
    pages = "035025",
    year = "2017"
}

@article{Okada:2018ktp,
    author = "Okada, Satomi",
    title = "{$Z'$ Portal Dark Matter in the Minimal $B-L$ Model}",
    eprint = "1803.06793",
    archivePrefix = "arXiv",
    primaryClass = "hep-ph",
    doi = "10.1155/2018/5340935",
    journal = "Adv. High Energy Phys.",
    volume = "2018",
    pages = "5340935",
    year = "2018"
}

@article{Okada:2016gsh,
    author = "Okada, Nobuchika and Okada, Satomi",
    title = "{$Z^\prime_{BL}$ portal dark matter and LHC Run-2 results}",
    eprint = "1601.07526",
    archivePrefix = "arXiv",
    primaryClass = "hep-ph",
    reportNumber = "YGHP16-03",
    doi = "10.1103/PhysRevD.93.075003",
    journal = "Phys. Rev. D",
    volume = "93",
    number = "7",
    pages = "075003",
    year = "2016"
}

@article{Wang:2015saa,
    author = "Wang, Weijian and Han, Zhi-Long",
    title = "{Radiative linear seesaw model, dark matter, and $U(1)_{B-L}$}",
    eprint = "1508.00706",
    archivePrefix = "arXiv",
    primaryClass = "hep-ph",
    doi = "10.1103/PhysRevD.92.095001",
    journal = "Phys. Rev. D",
    volume = "92",
    pages = "095001",
    year = "2015"
}

@article{Klasen:2016qux,
    author = "Klasen, Michael and Lyonnet, Florian and Queiroz, Farinaldo S.",
    title = "{NLO+NLL collider bounds, Dirac fermion and scalar dark matter in the B{\textendash}L model}",
    eprint = "1607.06468",
    archivePrefix = "arXiv",
    primaryClass = "hep-ph",
    reportNumber = "MITP-16-065",
    doi = "10.1140/epjc/s10052-017-4904-8",
    journal = "Eur. Phys. J. C",
    volume = "77",
    number = "5",
    pages = "348",
    year = "2017"
}

@article{Eijima:2022dec,
    author = "Eijima, Shintaro and Seto, Osamu and Shimomura, Takashi",
    title = "{Revisiting sterile neutrino dark matter in gauged U(1)B-L model}",
    eprint = "2207.01775",
    archivePrefix = "arXiv",
    primaryClass = "hep-ph",
    reportNumber = "UME-PP-021, EPHOU-22-013, KYUSHU-HET-244",
    doi = "10.1103/PhysRevD.106.103513",
    journal = "Phys. Rev. D",
    volume = "106",
    number = "10",
    pages = "103513",
    year = "2022"
}

@article{ACT:2025tim,
    author = "Calabrese, Erminia and others",
    collaboration = "ACT",
    title = "{The Atacama Cosmology Telescope: DR6 Constraints on Extended Cosmological Models}",
    eprint = "2503.14454",
    archivePrefix = "arXiv",
    primaryClass = "astro-ph.CO",
    reportNumber = "FERMILAB-PUB-25-0157-PPD",
    month = "3",
    year = "2025"
}

@article{Lazarides:1980nt,
    author = "Lazarides, George and Shafi, Q. and Wetterich, C.",
    title = "{Proton Lifetime and Fermion Masses in an SO(10) Model}",
    reportNumber = "FREIBURG-THEP-80-2",
    doi = "10.1016/0550-3213(81)90354-0",
    journal = "Nucl. Phys. B",
    volume = "181",
    pages = "287--300",
    year = "1981"
}

@article{Malinsky:2005bi,
	title        = {{Novel supersymmetric SO(10) seesaw mechanism}},
	author       = {Malinsky, Michal and Romao, J. C. and Valle, J. W. F.},
	year         = 2005,
	journal      = {Phys. Rev. Lett.},
	volume       = 95,
	pages        = 161801,
	doi          = {10.1103/PhysRevLett.95.161801},
	eprint       = {hep-ph/0506296},
	archiveprefix = {arXiv},
	primaryclass = {hep-ph},
	reportnumber = {IFIC-05-28},
	slaccitation = {%%CITATION = HEP-PH/0506296;%%}
}

@article{Minkowski:1977sc,
    author = "Minkowski, Peter",
    title = "{$\mu \to e\gamma$ at a Rate of One Out of $10^{9}$ Muon Decays?}",
    reportNumber = "Print-77-0182 (BERN)",
    doi = "10.1016/0370-2693(77)90435-X",
    journal = "Phys. Lett. B",
    volume = "67",
    pages = "421--428",
    year = "1977"
}

@article{Tulin:2017ara,
    author = "Tulin, Sean and Yu, Hai-Bo",
    title = "{Dark Matter Self-interactions and Small Scale Structure}",
    eprint = "1705.02358",
    archivePrefix = "arXiv",
    primaryClass = "hep-ph",
    doi = "10.1016/j.physrep.2017.11.004",
    journal = "Phys. Rept.",
    volume = "730",
    pages = "1--57",
    year = "2018"
}

@article{Adhikari:2022sbh,
    author = "Adhikari, Susmita and others",
    title = "{Astrophysical tests of dark matter self-interactions}",
    eprint = "2207.10638",
    archivePrefix = "arXiv",
    primaryClass = "astro-ph.CO",
    doi = "10.1103/m2vm-59y3",
    journal = "Rev. Mod. Phys.",
    volume = "97",
    number = "4",
    pages = "045004",
    year = "2025"
}

@article{Blennow:2013jba,
    author = "Blennow, Mattias and Fernandez-Martinez, Enrique and Zaldivar, Bryan",
    title = "{Freeze-in through portals}",
    eprint = "1309.7348",
    archivePrefix = "arXiv",
    primaryClass = "hep-ph",
    doi = "10.1088/1475-7516/2014/01/003",
    journal = "JCAP",
    volume = "01",
    pages = "003",
    year = "2014"
}

@article{Mohapatra:1979ia,
    author = "Mohapatra, Rabindra N. and Senjanovic, Goran",
    title = "{Neutrino Mass and Spontaneous Parity Nonconservation}",
    reportNumber = "MDDP-TR-80-060, MDDP-PP-80-105, CCNY-HEP-79-10",
    doi = "10.1103/PhysRevLett.44.912",
    journal = "Phys. Rev. Lett.",
    volume = "44",
    pages = "912",
    year = "1980"
}

@article{Yanagida:1979as,
    author = "Yanagida, Tsutomu",
    editor = "Sawada, Osamu and Sugamoto, Akio",
    title = "{Horizontal gauge symmetry and masses of neutrinos}",
    reportNumber = "KEK-79-18-95",
    journal = "Conf. Proc. C",
    volume = "7902131",
    pages = "95--99",
    year = "1979"
}

@article{Gell-Mann:1979vob,
    author = "Gell-Mann, Murray and Ramond, Pierre and Slansky, Richard",
    title = "{Complex Spinors and Unified Theories}",
    eprint = "1306.4669",
    archivePrefix = "arXiv",
    primaryClass = "hep-th",
    reportNumber = "PRINT-80-0576",
    journal = "Conf. Proc. C",
    volume = "790927",
    pages = "315--321",
    year = "1979"
}

@article{Yeh:2022heq,
    author = "Yeh, Tsung-Han and Shelton, Jessie and Olive, Keith A. and Fields, Brian D.",
    title = "{Probing physics beyond the standard model: limits from BBN and the CMB independently and combined}",
    eprint = "2207.13133",
    archivePrefix = "arXiv",
    primaryClass = "astro-ph.CO",
    reportNumber = "UMN-TH-4125/22, FTPI-MINN-22/16",
    doi = "10.1088/1475-7516/2022/10/046",
    journal = "JCAP",
    volume = "10",
    pages = "046",
    year = "2022"
}

@article{Fernandez-Martinez:2023phj,
    author = "Fern\'andez-Mart\'\i{}nez, Enrique and Gonz\'alez-L\'opez, Manuel and Hern\'andez-Garc\'\i{}a, Josu and Hostert, Matheus and L\'opez-Pav\'on, Jacobo",
    title = "{Effective portals to heavy neutral leptons}",
    eprint = "2304.06772",
    archivePrefix = "arXiv",
    primaryClass = "hep-ph",
    reportNumber = "FTUV-23-0303.1224, IFIC/23-09",
    doi = "10.1007/JHEP09(2023)001",
    journal = "JHEP",
    volume = "09",
    pages = "001",
    year = "2023"
}

@article{Davidson:2002qv,
    author = "Davidson, Sacha and Ibarra, Alejandro",
    title = "{A Lower bound on the right-handed neutrino mass from leptogenesis}",
    eprint = "hep-ph/0202239",
    archivePrefix = "arXiv",
    reportNumber = "OUTP-02-10P, IPPP-02-16, DCPT-02-32",
    doi = "10.1016/S0370-2693(02)01735-5",
    journal = "Phys. Lett. B",
    volume = "535",
    pages = "25--32",
    year = "2002"
}

@article{Drewes:2017zyw,
    author = "Drewes, M. and Garbrecht, B. and Hernandez, P. and Kekic, M. and Lopez-Pavon, J. and Racker, J. and Rius, N. and Salvado, J. and Teresi, D.",
    title = "{ARS Leptogenesis}",
    eprint = "1711.02862",
    archivePrefix = "arXiv",
    primaryClass = "hep-ph",
    doi = "10.1142/S0217751X18420022",
    journal = "Int. J. Mod. Phys. A",
    volume = "33",
    number = "05n06",
    pages = "1842002",
    year = "2018"
}

@article{Sabti:2020yrt,
    author = "Sabti, Nashwan and Magalich, Andrii and Filimonova, Anastasiia",
    title = "{An Extended Analysis of Heavy Neutral Leptons during Big Bang Nucleosynthesis}",
    eprint = "2006.07387",
    archivePrefix = "arXiv",
    primaryClass = "hep-ph",
    reportNumber = "KCL-2020-09",
    doi = "10.1088/1475-7516/2020/11/056",
    journal = "JCAP",
    volume = "11",
    pages = "056",
    year = "2020"
}

@article{Boyarsky:2020dzc,
    author = "Boyarsky, Alexey and Ovchynnikov, Maksym and Ruchayskiy, Oleg and Syvolap, Vsevolod",
    title = "{Improved big bang nucleosynthesis constraints on heavy neutral leptons}",
    eprint = "2008.00749",
    archivePrefix = "arXiv",
    primaryClass = "hep-ph",
    doi = "10.1103/PhysRevD.104.023517",
    journal = "Phys. Rev. D",
    volume = "104",
    number = "2",
    pages = "023517",
    year = "2021"
}

@article{Dolgov:2000jw,
    author = "Dolgov, A. D. and Hansen, S. H. and Raffelt, G. and Semikoz, D. V.",
    title = "{Heavy sterile neutrinos: Bounds from big bang nucleosynthesis and SN1987A}",
    eprint = "hep-ph/0008138",
    archivePrefix = "arXiv",
    doi = "10.1016/S0550-3213(00)00566-6",
    journal = "Nucl. Phys. B",
    volume = "590",
    pages = "562--574",
    year = "2000"
}

@article{Akhmedov:1998qx,
    author = "Akhmedov, Evgeny K. and Rubakov, V. A. and Smirnov, A. Yu.",
    title = "{Baryogenesis via neutrino oscillations}",
    eprint = "hep-ph/9803255",
    archivePrefix = "arXiv",
    reportNumber = "IC-98-22, INR-98-14-T",
    doi = "10.1103/PhysRevLett.81.1359",
    journal = "Phys. Rev. Lett.",
    volume = "81",
    pages = "1359--1362",
    year = "1998"
}

@article{Asaka:2005pn,
    author = "Asaka, Takehiko and Shaposhnikov, Mikhail",
    title = "{The $\nu$MSM, dark matter and baryon asymmetry of the universe}",
    eprint = "hep-ph/0505013",
    archivePrefix = "arXiv",
    doi = "10.1016/j.physletb.2005.06.020",
    journal = "Phys. Lett. B",
    volume = "620",
    pages = "17--26",
    year = "2005"
}

@article{Caputo:2018zky,
    author = "Caputo, Andrea and Hernandez, Pilar and Rius, Nuria",
    title = "{Leptogenesis from oscillations and dark matter}",
    eprint = "1807.03309",
    archivePrefix = "arXiv",
    primaryClass = "hep-ph",
    doi = "10.1140/epjc/s10052-019-7083-y",
    journal = "Eur. Phys. J. C",
    volume = "79",
    number = "7",
    pages = "574",
    year = "2019"
}

@article{Abada:2014vea,
    author = "Abada, Asmaa and Lucente, Michele",
    title = "{Looking for the minimal inverse seesaw realisation}",
    eprint = "1401.1507",
    archivePrefix = "arXiv",
    primaryClass = "hep-ph",
    reportNumber = "LPT-ORSAY-13-39, SISSA-59-2013-FISI",
    doi = "10.1016/j.nuclphysb.2014.06.003",
    journal = "Nucl. Phys. B",
    volume = "885",
    pages = "651--678",
    year = "2014"
}

@article{LZ:2018qzl,
    author = "Akerib, D. S. and others",
    collaboration = "LZ",
    title = "{Projected WIMP sensitivity of the LUX-ZEPLIN dark matter experiment}",
    eprint = "1802.06039",
    archivePrefix = "arXiv",
    primaryClass = "astro-ph.IM",
    reportNumber = "FERMILAB-PUB-18-054-AE-PPD",
    doi = "10.1103/PhysRevD.101.052002",
    journal = "Phys. Rev. D",
    volume = "101",
    number = "5",
    pages = "052002",
    year = "2020"
}

@article{Fitzpatrick:2012ix,
    author = "Fitzpatrick, A. Liam and Haxton, Wick and Katz, Emanuel and Lubbers, Nicholas and Xu, Yiming",
    title = "{The Effective Field Theory of Dark Matter Direct Detection}",
    eprint = "1203.3542",
    archivePrefix = "arXiv",
    primaryClass = "hep-ph",
    doi = "10.1088/1475-7516/2013/02/004",
    journal = "JCAP",
    volume = "02",
    pages = "004",
    year = "2013"
}

@article{Anand:2013yka,
    author = "Anand, Nikhil and Fitzpatrick, A. Liam and Haxton, W. C.",
    title = "{Weakly interacting massive particle-nucleus elastic scattering response}",
    eprint = "1308.6288",
    archivePrefix = "arXiv",
    primaryClass = "hep-ph",
    doi = "10.1103/PhysRevC.89.065501",
    journal = "Phys. Rev. C",
    volume = "89",
    number = "6",
    pages = "065501",
    year = "2014"
}

@article{DelNobile:2021wmp,
    author = "Del Nobile, Eugenio",
    title = "{The Theory of Direct Dark Matter Detection: A Guide to Computations}",
    eprint = "2104.12785",
    archivePrefix = "arXiv",
    primaryClass = "hep-ph",
    doi = "10.1007/978-3-030-95228-0",
    month = "4",
    year = "2021"
}

@article{Baudis:2024jnk,
    author = "Baudis, Laura",
    title = "{DARWIN/XLZD: A future xenon observatory for dark matter and other rare interactions}",
    eprint = "2404.19524",
    archivePrefix = "arXiv",
    primaryClass = "astro-ph.IM",
    doi = "10.1016/j.nuclphysb.2024.116473",
    journal = "Nucl. Phys. B",
    volume = "1003",
    pages = "116473",
    year = "2024"
}

@article{Alekhin:2015byh,
    author = "Alekhin, Sergey and others",
    title = "{A facility to Search for Hidden Particles at the CERN SPS: the SHiP physics case}",
    eprint = "1504.04855",
    archivePrefix = "arXiv",
    primaryClass = "hep-ph",
    reportNumber = "CERN-SPSC-2015-017, SPSC-P-350-ADD-1",
    doi = "10.1088/0034-4885/79/12/124201",
    journal = "Rept. Prog. Phys.",
    volume = "79",
    number = "12",
    pages = "124201",
    year = "2016"
}

@article{BaBar:2017tiz,
    author = "Lees, J. P. and others",
    collaboration = "BaBar",
    title = "{Search for Invisible Decays of a Dark Photon Produced in ${e}^{+}{e}^{-}$ Collisions at BaBar}",
    eprint = "1702.03327",
    archivePrefix = "arXiv",
    primaryClass = "hep-ex",
    reportNumber = "BABAR-PUB-17-001, SLAC-PUB-16923",
    doi = "10.1103/PhysRevLett.119.131804",
    journal = "Phys. Rev. Lett.",
    volume = "119",
    number = "13",
    pages = "131804",
    year = "2017"
}

@article{Cesarotti:2022ttv,
    author = "Cesarotti, Cari and Homiller, Samuel and Mishra, Rashmish K. and Reece, Matthew",
    title = "{Probing New Gauge Forces with a High-Energy Muon Beam Dump}",
    eprint = "2202.12302",
    archivePrefix = "arXiv",
    primaryClass = "hep-ph",
    doi = "10.1103/PhysRevLett.130.071803",
    journal = "Phys. Rev. Lett.",
    volume = "130",
    number = "7",
    pages = "071803",
    year = "2023"
}

@inproceedings{Apyan:2022tsd,
    author = "Apyan, Aram and others",
    title = "{DarkQuest: A dark sector upgrade to SpinQuest at the 120 GeV Fermilab Main Injector}",
    booktitle = "{Snowmass 2021}",
    eprint = "2203.08322",
    archivePrefix = "arXiv",
    primaryClass = "hep-ex",
    reportNumber = "FERMILAB-CONF-22-175-PPD-SCD-T",
    month = "3",
    year = "2022"
}

@article{Airen:2024iiy,
    author = "Airen, Sagar and Broadberry, Edward and Marques-Tavares, Gustavo and Ricci, Lorenzo",
    title = "{Vector Portals at Future Lepton Colliders}",
    eprint = "2412.09681",
    archivePrefix = "arXiv",
    primaryClass = "hep-ph",
    month = "12",
    year = "2024"
}

@article{Belle-II:2018jsg,
    author = "Altmannshofer, W. and others",
    editor = "Kou, E. and Urquijo, P.",
    collaboration = "Belle-II",
    title = "{The Belle II Physics Book}",
    eprint = "1808.10567",
    archivePrefix = "arXiv",
    primaryClass = "hep-ex",
    reportNumber = "KEK Preprint 2018-27, BELLE2-PUB-PH-2018-001, FERMILAB-PUB-18-398-T, JLAB-THY-18-2780, INT-PUB-18-047, UWThPh 2018-26",
    doi = "10.1093/ptep/ptz106",
    journal = "PTEP",
    volume = "2019",
    number = "12",
    pages = "123C01",
    year = "2019",
    note = "[Erratum: PTEP 2020, 029201 (2020)]"
}

@article{Belle-II:2022cgf,
    author = "Aggarwal, Latika and others",
    collaboration = "Belle-II",
    title = "{Snowmass White Paper: Belle II physics reach and plans for the next decade and beyond}",
    eprint = "2207.06307",
    archivePrefix = "arXiv",
    primaryClass = "hep-ex",
    month = "7",
    year = "2022"
}

@article{NA64:2016oww,
    author = "Banerjee, D. and others",
    collaboration = "NA64",
    title = "{Search for invisible decays of sub-GeV dark photons in missing-energy events at the CERN SPS}",
    eprint = "1610.02988",
    archivePrefix = "arXiv",
    primaryClass = "hep-ex",
    doi = "10.1103/PhysRevLett.118.011802",
    journal = "Phys. Rev. Lett.",
    volume = "118",
    number = "1",
    pages = "011802",
    year = "2017"
}

@article{NA64:2024nwj,
    author = "Andreev, Yu. M. and others",
    collaboration = "NA64",
    title = "{Shedding light on dark sectors with high-energy muons at the NA64 experiment at the CERN SPS}",
    eprint = "2409.10128",
    archivePrefix = "arXiv",
    primaryClass = "hep-ex",
    reportNumber = "CERN-EP-2024-236",
    doi = "10.1103/PhysRevD.110.112015",
    journal = "Phys. Rev. D",
    volume = "110",
    number = "11",
    pages = "112015",
    year = "2024"
}

@article{Cielo:2023bqp,
    author = "Cielo, Mattia and Escudero, Miguel and Mangano, Gianpiero and Pisanti, Ofelia",
    title = "{Neff in the Standard Model at NLO is 3.043}",
    eprint = "2306.05460",
    archivePrefix = "arXiv",
    primaryClass = "hep-ph",
    reportNumber = "CERN-TH-2023-103",
    doi = "10.1103/PhysRevD.108.L121301",
    journal = "Phys. Rev. D",
    volume = "108",
    number = "12",
    pages = "L121301",
    year = "2023"
}

@article{Escudero:2025kej,
    author = "Escudero, M. and Jackson, G. and Laine, M. and Sandner, S.",
    title = "{Fast and Flexible Neutrino Decoupling Part I: The Standard Model}",
    eprint = "2511.04747",
    archivePrefix = "arXiv",
    primaryClass = "hep-ph",
    reportNumber = "LA-UR-25-30442, CERN-TH-2025-225",
    month = "11",
    year = "2025"
}

@article{Planck:2018vyg,
    author = "Aghanim, N. and others",
    collaboration = "Planck",
    title = "{Planck 2018 results. VI. Cosmological parameters}",
    eprint = "1807.06209",
    archivePrefix = "arXiv",
    primaryClass = "astro-ph.CO",
    doi = "10.1051/0004-6361/201833910",
    journal = "Astron. Astrophys.",
    volume = "641",
    pages = "A6",
    year = "2020",
    note = "[Erratum: Astron.Astrophys. 652, C4 (2021)]"
}

@article{Ovchynnikov:2024xyd,
    author = "Ovchynnikov, Maksym and Syvolap, Vsevolod",
    title = "{Primordial Neutrinos and New Physics: Novel Approach to Solving the Neutrino Boltzmann Equation}",
    eprint = "2409.15129",
    archivePrefix = "arXiv",
    primaryClass = "hep-ph",
    reportNumber = "CERN-TH-2024-159",
    doi = "10.1103/PhysRevLett.134.101003",
    journal = "Phys. Rev. Lett.",
    volume = "134",
    number = "10",
    pages = "101003",
    year = "2025"
}

@article{SimonsObservatory:2025wwn,
    author = "Abitbol, M. and others",
    collaboration = "Simons Observatory",
    title = "{The Simons Observatory: science goals and forecasts for the enhanced Large Aperture Telescope}",
    eprint = "2503.00636",
    archivePrefix = "arXiv",
    primaryClass = "astro-ph.IM",
    reportNumber = "FERMILAB-PUB-25-0188-PPD",
    doi = "10.1088/1475-7516/2025/08/034",
    journal = "JCAP",
    volume = "08",
    pages = "034",
    year = "2025"
}

@article{Ilten:2018crw,
    author = "Ilten, Philip and Soreq, Yotam and Williams, Mike and Xue, Wei",
    title = "{Serendipity in dark photon searches}",
    eprint = "1801.04847",
    archivePrefix = "arXiv",
    primaryClass = "hep-ph",
    reportNumber = "MIT-CTP/4976, CERN-TH-2017-282, MIT-CTP-4976",
    doi = "10.1007/JHEP06(2018)004",
    journal = "JHEP",
    volume = "06",
    pages = "004",
    year = "2018"
}

@article{Foguel:2024lca,
    author = "Foguel, Ana Luisa and Reimitz, Peter and Funchal, Renata Zukanovich",
    title = "{Unlocking the inelastic Dark Matter window with vector mediators}",
    eprint = "2410.00881",
    archivePrefix = "arXiv",
    primaryClass = "hep-ph",
    doi = "10.1007/JHEP05(2025)001",
    journal = "JHEP",
    volume = "05",
    pages = "001",
    year = "2025"
}

@article{Kling:2021fwx,
    author = "Kling, Felix and Trojanowski, Sebastian",
    title = "{Forward experiment sensitivity estimator for the LHC and future hadron colliders}",
    eprint = "2105.07077",
    archivePrefix = "arXiv",
    primaryClass = "hep-ph",
    doi = "10.1103/PhysRevD.104.035012",
    journal = "Phys. Rev. D",
    volume = "104",
    number = "3",
    pages = "035012",
    year = "2021"
}

@article{Ovchynnikov:2023cry,
    author = "Ovchynnikov, Maksym and Tastet, Jean-Loup and Mikulenko, Oleksii and Bondarenko, Kyrylo",
    title = "{Sensitivities to feebly interacting particles: Public and unified calculations}",
    eprint = "2305.13383",
    archivePrefix = "arXiv",
    primaryClass = "hep-ph",
    doi = "10.1103/PhysRevD.108.075028",
    journal = "Phys. Rev. D",
    volume = "108",
    number = "7",
    pages = "075028",
    year = "2023"
}

@article{Foguel:2022ppx,
    author = "Foguel, Ana Luisa and Reimitz, Peter and Funchal, Renata Zukanovich",
    title = "{A robust description of hadronic decays in light vector mediator models}",
    eprint = "2201.01788",
    archivePrefix = "arXiv",
    primaryClass = "hep-ph",
    doi = "10.1007/JHEP04(2022)119",
    journal = "JHEP",
    volume = "04",
    pages = "119",
    year = "2022"
}

@article{Bjorken:1988as,
    author = "Bjorken, J. D. and Ecklund, S. and Nelson, W. R. and Abashian, A. and Church, C. and Lu, B. and Mo, L. W. and Nunamaker, T. A. and Rassmann, P.",
    title = "{Search for Neutral Metastable Penetrating Particles Produced in the SLAC Beam Dump}",
    reportNumber = "FERMILAB-PUB-88-044, PRINT-88-0352 (FERMILAB)",
    doi = "10.1103/PhysRevD.38.3375",
    journal = "Phys. Rev. D",
    volume = "38",
    pages = "3375",
    year = "1988"
}

@article{Blumlein:2013cua,
    author = {Bl{\"u}mlein, Johannes and Brunner, J{\"u}rgen},
    title = "{New Exclusion Limits on Dark Gauge Forces from Proton Bremsstrahlung in Beam-Dump Data}",
    eprint = "1311.3870",
    archivePrefix = "arXiv",
    primaryClass = "hep-ph",
    reportNumber = "DESY-13-202, DO-TH-13-29, SFB-CPP-13-87, LPN-13-087",
    doi = "10.1016/j.physletb.2014.02.029",
    journal = "Phys. Lett. B",
    volume = "731",
    pages = "320--326",
    year = "2014"
}

@article{Blumlein:2011mv,
    author = "Blumlein, Johannes and Brunner, Jurgen",
    title = "{New Exclusion Limits for Dark Gauge Forces from Beam-Dump Data}",
    eprint = "1104.2747",
    archivePrefix = "arXiv",
    primaryClass = "hep-ex",
    reportNumber = "DESY-11-062, DO-TH-11-11, SFB-CPP-11-18, LPN-11-17, DESY-11--062, DO--TH-11-11, SFB-CPP--11--18, LPN-11--17",
    doi = "10.1016/j.physletb.2011.05.046",
    journal = "Phys. Lett. B",
    volume = "701",
    pages = "155--159",
    year = "2011"
}

@article{CHARM:1985anb,
    author = "Bergsma, F. and others",
    collaboration = "CHARM",
    title = "{Search for Axion Like Particle Production in 400-{GeV} Proton - Copper Interactions}",
    reportNumber = "CERN-EP-85-38",
    doi = "10.1016/0370-2693(85)90400-9",
    journal = "Phys. Lett. B",
    volume = "157",
    pages = "458--462",
    year = "1985"
}

@article{Foroughi-Abari:2024xlj,
    author = "Foroughi-Abari, Saeid and Reimitz, Peter and Ritz, Adam",
    title = "{Closer look at dark vector splitting functions in proton bremsstrahlung}",
    eprint = "2409.09123",
    archivePrefix = "arXiv",
    primaryClass = "hep-ph",
    doi = "10.1103/fzlm-gsd7",
    journal = "Phys. Rev. D",
    volume = "112",
    number = "1",
    pages = "015030",
    year = "2025"
}

@article{Gorbunov:2014wqa,
    author = "Gorbunov, D. and Makarov, A. and Timiryasov, I.",
    title = "{Decaying light particles in the SHiP experiment: Signal rate estimates for hidden photons}",
    eprint = "1411.4007",
    archivePrefix = "arXiv",
    primaryClass = "hep-ph",
    reportNumber = "INR-TH-2014-026, INR-TH/2014-026",
    doi = "10.1103/PhysRevD.91.035027",
    journal = "Phys. Rev. D",
    volume = "91",
    number = "3",
    pages = "035027",
    year = "2015"
}

@article{Carloni:2011kk,
    author = "Carloni, Lisa and Rathsman, Johan and Sjostrand, Torbjorn",
    title = "{Discerning Secluded Sector gauge structures}",
    eprint = "1102.3795",
    archivePrefix = "arXiv",
    primaryClass = "hep-ph",
    reportNumber = "LU-TP-11-09, MCNET-11-06",
    doi = "10.1007/JHEP04(2011)091",
    journal = "JHEP",
    volume = "04",
    pages = "091",
    year = "2011"
}

@article{SHiP:2021nfo,
    author = "Ahdida, C. and others",
    collaboration = "SHiP",
    title = "{The SHiP experiment at the proposed CERN SPS Beam Dump Facility}",
    eprint = "2112.01487",
    archivePrefix = "arXiv",
    primaryClass = "physics.ins-det",
    doi = "10.1140/epjc/s10052-022-10346-5",
    journal = "Eur. Phys. J. C",
    volume = "82",
    number = "5",
    pages = "486",
    year = "2022"
}

@article{Bjorken:2009mm,
    author = "Bjorken, James D. and Essig, Rouven and Schuster, Philip and Toro, Natalia",
    title = "{New Fixed-Target Experiments to Search for Dark Gauge Forces}",
    eprint = "0906.0580",
    archivePrefix = "arXiv",
    primaryClass = "hep-ph",
    reportNumber = "SLAC-PUB-13650, SU-ITP-09-22",
    doi = "10.1103/PhysRevD.80.075018",
    journal = "Phys. Rev. D",
    volume = "80",
    pages = "075018",
    year = "2009"
}

@article{Kirpichnikov:2021jev,
    author = "Kirpichnikov, D. V. and Sieber, H. and Bueno, L. Molina and Crivelli, P. and Kirsanov, M. M.",
    title = "{Probing hidden sectors with a muon beam: Total and differential cross sections for vector boson production in muon bremsstrahlung}",
    eprint = "2107.13297",
    archivePrefix = "arXiv",
    primaryClass = "hep-ph",
    doi = "10.1103/PhysRevD.104.076012",
    journal = "Phys. Rev. D",
    volume = "104",
    number = "7",
    pages = "076012",
    year = "2021"
}

@article{Liu:2017htz,
    author = "Liu, Yu-Sheng and Miller, Gerald A.",
    title = {{Validity of the Weizs{\"a}cker-Williams approximation and the analysis of beam dump experiments: Production of an axion, a dark photon, or a new axial-vector boson}},
    eprint = "1705.01633",
    archivePrefix = "arXiv",
    primaryClass = "hep-ph",
    reportNumber = "NT@UW-17-05",
    doi = "10.1103/PhysRevD.96.016004",
    journal = "Phys. Rev. D",
    volume = "96",
    number = "1",
    pages = "016004",
    year = "2017"
}

@article{Batell:2009di,
    author = "Batell, Brian and Pospelov, Maxim and Ritz, Adam",
    title = "{Exploring Portals to a Hidden Sector Through Fixed Targets}",
    eprint = "0906.5614",
    archivePrefix = "arXiv",
    primaryClass = "hep-ph",
    doi = "10.1103/PhysRevD.80.095024",
    journal = "Phys. Rev. D",
    volume = "80",
    pages = "095024",
    year = "2009"
}

@article{SPT-3G:2025bzu,
    author = "Camphuis, E. and others",
    collaboration = "SPT-3G",
    title = "{SPT-3G D1: CMB temperature and polarization power spectra and cosmology from 2019 and 2020 observations of the SPT-3G Main field}",
    eprint = "2506.20707",
    archivePrefix = "arXiv",
    primaryClass = "astro-ph.CO",
    reportNumber = "FERMILAB-PUB-25-0144-PPD",
    month = "6",
    year = "2025"
}

@article{DEramo:2024jhn,
    author = "D'Eramo, Francesco and Lenoci, Alessandro",
    title = "{Back to the phase space: Thermal axion dark radiation via couplings to standard model fermions}",
    eprint = "2410.21253",
    archivePrefix = "arXiv",
    primaryClass = "hep-ph",
    doi = "10.1103/PhysRevD.110.116028",
    journal = "Phys. Rev. D",
    volume = "110",
    number = "11",
    pages = "116028",
    year = "2024"
}

@article{Klaric:2021cpi,
    author = "Klari{\'c}, Juraj and Shaposhnikov, Mikhail and Timiryasov, Inar",
    title = "{Reconciling resonant leptogenesis and baryogenesis via neutrino oscillations}",
    eprint = "2103.16545",
    archivePrefix = "arXiv",
    primaryClass = "hep-ph",
    doi = "10.1103/PhysRevD.104.055010",
    journal = "Phys. Rev. D",
    volume = "104",
    number = "5",
    pages = "055010",
    year = "2021"
}

@article{Green:2019glg,
    author = "Green, Daniel and others",
    title = "{Messengers from the Early Universe: Cosmic Neutrinos and Other Light Relics}",
    eprint = "1903.04763",
    archivePrefix = "arXiv",
    primaryClass = "astro-ph.CO",
    reportNumber = "FERMILAB-PUB-19-099-A-AE-CD",
    journal = "Bull. Am. Astron. Soc.",
    volume = "51",
    number = "3",
    pages = "159",
    year = "2019"
}

@article{Riordan:1987aw,
    author = "Riordan, E. M. and others",
    title = "{A Search for Short Lived Axions in an Electron Beam Dump Experiment}",
    reportNumber = "SLAC-PUB-4280, UR-993, FERMILAB-PUB-87-251",
    doi = "10.1103/PhysRevLett.59.755",
    journal = "Phys. Rev. Lett.",
    volume = "59",
    pages = "755",
    year = "1987"
}

@article{Cesarotti:2023sje,
    author = "Cesarotti, Cari and Gambhir, Rikab",
    title = "{The new physics case for beam-dump experiments with accelerated muon beams}",
    eprint = "2310.16110",
    archivePrefix = "arXiv",
    primaryClass = "hep-ph",
    reportNumber = "MIT-CTP 5606",
    doi = "10.1007/JHEP05(2024)283",
    journal = "JHEP",
    volume = "05",
    pages = "283",
    year = "2024"
}

@article{BaBar:2014zli,
    author = "Lees, J. P. and others",
    collaboration = "BaBar",
    title = "{Search for a Dark Photon in $e^+e^-$ Collisions at BaBar}",
    eprint = "1406.2980",
    archivePrefix = "arXiv",
    primaryClass = "hep-ex",
    reportNumber = "BABAR-PUB-14-002, SLAC-PUB-15979",
    doi = "10.1103/PhysRevLett.113.201801",
    journal = "Phys. Rev. Lett.",
    volume = "113",
    number = "20",
    pages = "201801",
    year = "2014"
}

@article{Fox:2011fx,
    author = "Fox, Patrick J. and Harnik, Roni and Kopp, Joachim and Tsai, Yuhsin",
    title = "{LEP Shines Light on Dark Matter}",
    eprint = "1103.0240",
    archivePrefix = "arXiv",
    primaryClass = "hep-ph",
    reportNumber = "FERMILAB-PUB-11-039-T",
    doi = "10.1103/PhysRevD.84.014028",
    journal = "Phys. Rev. D",
    volume = "84",
    pages = "014028",
    year = "2011"
}

@article{Zhang:2019wnz,
    author = "Zhang, Yu and Zhang, Wei-Tao and Song, Mao and Pan, Xue-An and Niu, Zhong-Ming and Li, Gang",
    title = "{Probing invisible decay of dark photon at BESIII and future STCF via monophoton searches}",
    eprint = "1907.07046",
    archivePrefix = "arXiv",
    primaryClass = "hep-ph",
    doi = "10.1103/PhysRevD.100.115016",
    journal = "Phys. Rev. D",
    volume = "100",
    number = "11",
    pages = "115016",
    year = "2019"
}

@article{LHCb:2017trq,
    author = "Aaij, Roel and others",
    collaboration = "LHCb",
    title = "{Search for Dark Photons Produced in 13 TeV $pp$ Collisions}",
    eprint = "1710.02867",
    archivePrefix = "arXiv",
    primaryClass = "hep-ex",
    reportNumber = "LHCB-PAPER-2017-038, CERN-EP-2017-248",
    doi = "10.1103/PhysRevLett.120.061801",
    journal = "Phys. Rev. Lett.",
    volume = "120",
    number = "6",
    pages = "061801",
    year = "2018"
}

@article{CMS:2019buh,
    author = "Sirunyan, Albert M and others",
    collaboration = "CMS",
    title = "{Search for a Narrow Resonance Lighter than 200 GeV Decaying to a Pair of Muons in Proton-Proton Collisions at $\sqrt{s} =$  TeV}",
    eprint = "1912.04776",
    archivePrefix = "arXiv",
    primaryClass = "hep-ex",
    reportNumber = "CMS-EXO-19-018, CERN-EP-2019-265",
    doi = "10.1103/PhysRevLett.124.131802",
    journal = "Phys. Rev. Lett.",
    volume = "124",
    number = "13",
    pages = "131802",
    year = "2020"
}

@article{Ilten:2015hya,
    author = "Ilten, Philip and Thaler, Jesse and Williams, Mike and Xue, Wei",
    title = "{Dark photons from charm mesons at LHCb}",
    eprint = "1509.06765",
    archivePrefix = "arXiv",
    primaryClass = "hep-ph",
    reportNumber = "MIT-CTP-4702",
    doi = "10.1103/PhysRevD.92.115017",
    journal = "Phys. Rev. D",
    volume = "92",
    number = "11",
    pages = "115017",
    year = "2015"
}

@article{Ilten:2016tkc,
    author = "Ilten, Philip and Soreq, Yotam and Thaler, Jesse and Williams, Mike and Xue, Wei",
    title = "{Proposed Inclusive Dark Photon Search at LHCb}",
    eprint = "1603.08926",
    archivePrefix = "arXiv",
    primaryClass = "hep-ph",
    reportNumber = "MIT-CTP-4785",
    doi = "10.1103/PhysRevLett.116.251803",
    journal = "Phys. Rev. Lett.",
    volume = "116",
    number = "25",
    pages = "251803",
    year = "2016"
}

@article{Ferber:2022ewf,
    author = "Ferber, Torben and Garcia-Cely, Camilo and Schmidt-Hoberg, Kai",
    title = "{BelleII sensitivity to long{\textendash}lived dark photons}",
    eprint = "2202.03452",
    archivePrefix = "arXiv",
    primaryClass = "hep-ph",
    doi = "10.1016/j.physletb.2022.137373",
    journal = "Phys. Lett. B",
    volume = "833",
    pages = "137373",
    year = "2022"
}

@article{Jaeckel:2023huy,
    author = "Jaeckel, Joerg and Phan, Anh Vu",
    title = "{Searching dark photons using displaced vertices at Belle II {\textemdash} with backgrounds}",
    eprint = "2312.12522",
    archivePrefix = "arXiv",
    primaryClass = "hep-ph",
    doi = "10.1007/JHEP08(2024)062",
    journal = "JHEP",
    volume = "08",
    pages = "062",
    year = "2024"
}

@article{FASER:2018eoc,
    author = "Ariga, Akitaka and others",
    collaboration = "FASER",
    title = "{FASER{\textquoteright}s physics reach for long-lived particles}",
    eprint = "1811.12522",
    archivePrefix = "arXiv",
    primaryClass = "hep-ph",
    reportNumber = "UCI-TR-2018-19, KYUSHU-RCAPP-2018-06",
    doi = "10.1103/PhysRevD.99.095011",
    journal = "Phys. Rev. D",
    volume = "99",
    number = "9",
    pages = "095011",
    year = "2019"
}

@article{Blondel:2014bra,
    author = "Blondel, Alain and Graverini, E. and Serra, N. and Shaposhnikov, M.",
    editor = "Aguilar-Ben{\'\i}tez, M and Fuster, J and Mart{\'\i}-Garc{\'\i}a, S and Santamar{\'\i}a, A",
    collaboration = "FCC-ee study Team",
    title = "{Search for Heavy Right Handed Neutrinos at the FCC-ee}",
    eprint = "1411.5230",
    archivePrefix = "arXiv",
    primaryClass = "hep-ex",
    doi = "10.1016/j.nuclphysbps.2015.09.304",
    journal = "Nucl. Part. Phys. Proc.",
    volume = "273-275",
    pages = "1883--1890",
    year = "2016"
}

@article{Okada:2012np,
    author = "Okada, Hiroshi and Toma, Takashi",
    title = "{Fermionic Dark Matter in Radiative Inverse Seesaw Model with $U(1)_{B-L}$}",
    eprint = "1207.0864",
    archivePrefix = "arXiv",
    primaryClass = "hep-ph",
    doi = "10.1103/PhysRevD.86.033011",
    journal = "Phys. Rev. D",
    volume = "86",
    pages = "033011",
    year = "2012"
}

@article{Kajiyama:2012xg,
    author = "Kajiyama, Yuji and Okada, Hiroshi and Toma, Takashi",
    title = "{Light Dark Matter Candidate in B-L Gauged Radiative Inverse Seesaw}",
    eprint = "1210.2305",
    archivePrefix = "arXiv",
    primaryClass = "hep-ph",
    reportNumber = "KIAS-P12066, IPPP-12-75, DCPT-12-150",
    doi = "10.1140/epjc/s10052-013-2381-2",
    journal = "Eur. Phys. J. C",
    volume = "73",
    number = "3",
    pages = "2381",
    year = "2013"
}

@article{Seto:2024lik,
    author = "Seto, Osamu and Shimomura, Takashi and Uchida, Yoshiki",
    title = "{Freeze-in sterile neutrino dark matter in a feebly gauged B {\ensuremath{-}} L model}",
    eprint = "2404.00654",
    archivePrefix = "arXiv",
    primaryClass = "hep-ph",
    reportNumber = "UME-PP-027, EPHOU-24-003",
    doi = "10.1007/JHEP05(2025)147",
    journal = "JHEP",
    volume = "05",
    pages = "147",
    year = "2025"
}

@article{Giovanetti:2024eff,
    author = "Giovanetti, Cara and Lisanti, Mariangela and Liu, Hongwan and Mishra-Sharma, Siddharth and Ruderman, Joshua T.",
    title = "{Cosmological parameter estimation with a joint-likelihood analysis of the cosmic microwave background and big bang nucleosynthesis}",
    eprint = "2408.14531",
    archivePrefix = "arXiv",
    primaryClass = "astro-ph.CO",
    reportNumber = "FERMILAB-PUB-24-0738-V",
    doi = "10.1103/wspy-s948",
    journal = "Phys. Rev. D",
    volume = "112",
    number = "6",
    pages = "063530",
    year = "2025"
}

@article{Bringmann:2021sth,
    author = "Bringmann, Torsten and Heeba, Saniya and Kahlhoefer, Felix and Vangsnes, Kristian",
    title = "{Freezing-in a hot bath: resonances, medium effects and phase transitions}",
    eprint = "2111.14871",
    archivePrefix = "arXiv",
    primaryClass = "hep-ph",
    doi = "10.1007/JHEP02(2022)110",
    journal = "JHEP",
    volume = "02",
    pages = "110",
    year = "2022"
}

@article{Heeba:2019jho,
    author = "Heeba, Saniya and Kahlhoefer, Felix",
    title = "{Probing the freeze-in mechanism in dark matter models with U(1)' gauge extensions}",
    eprint = "1908.09834",
    archivePrefix = "arXiv",
    primaryClass = "hep-ph",
    reportNumber = "TTK-19-33",
    doi = "10.1103/PhysRevD.101.035043",
    journal = "Phys. Rev. D",
    volume = "101",
    number = "3",
    pages = "035043",
    year = "2020"
}

@article{Kaneta:2016vkq,
    author = "Kaneta, Kunio and Kang, Zhaofeng and Lee, Hye-Sung",
    title = "{Right-handed neutrino dark matter under the $B - L$ gauge interaction}",
    eprint = "1606.09317",
    archivePrefix = "arXiv",
    primaryClass = "hep-ph",
    reportNumber = "CTPU-16-17",
    doi = "10.1007/JHEP02(2017)031",
    journal = "JHEP",
    volume = "02",
    pages = "031",
    year = "2017"
}

@article{Bondarenko:2019yob,
    author = "Bondarenko, Kyrylo and Boyarsky, Alexey and Ovchynnikov, Maksym and Ruchayskiy, Oleg",
    title = "{Sensitivity of the intensity frontier experiments for neutrino and scalar portals: analytic estimates}",
    eprint = "1902.06240",
    archivePrefix = "arXiv",
    primaryClass = "hep-ph",
    doi = "10.1007/JHEP08(2019)061",
    journal = "JHEP",
    volume = "08",
    pages = "061",
    year = "2019"
}

@inproceedings{Arrenberg:2013rzp,
    author = "Arrenberg, Sebastian and others",
    title = "{Working Group Report: Dark Matter Complementarity}",
    booktitle = "{Snowmass 2013}: {Snowmass on the Mississippi}",
    eprint = "1310.8621",
    archivePrefix = "arXiv",
    primaryClass = "hep-ph",
    reportNumber = "UFIFT-HEP-13-001, FERMILAB-CONF-13-574-AE",
    month = "10",
    year = "2013"
}

\end{document}